\documentclass[12pt]{article}
\usepackage[a4paper, total={6in, 9in}]{geometry}
\usepackage[utf8]{inputenc}
\usepackage{natbib}
\usepackage{amsmath}
\usepackage{lineno}
\usepackage{setspace}
\usepackage{float}
\usepackage{tikz}
\usetikzlibrary{decorations.pathreplacing,positioning, arrows.meta, calc}
\usepackage{booktabs}
\usepackage{hyperref}
\hypersetup{
	colorlinks=true,
	linkcolor=blue,
	filecolor=magenta,      
	urlcolor=cyan,
	pdftitle={Overleaf Example},
	pdfpagemode=FullScreen,
}

\usepackage[english,hyperpageref]{backref}	 
\usepackage[alf]{abntex2cite}	

\usepackage[english]{babel}
\usepackage{pdflscape}
\usepackage{rotating}
\usepackage{rotfloat}
\usepackage{subfig}

\title{Is Crime Displacement Inevitable? Evidence from Police Crackdowns in Fortaleza, Brazil} 
\author{Jos{\'e} Raimundo Carvalho\thanks{Full Professor, CAEN/UFC Brazil \& Intelligence Analyst, Ministry of Defense.} \\CAEN/UFC \& Ministry of Defense.  \and Marcelino Guerra\thanks{Corresponding author: marcelinobguerra@gmail.com} \\ University of Illinois at Urbana-Champaign}

\date{\today}

		\makeatletter
		\hypersetup{
			pdftitle={\@title}, 
			pdfauthor={\@author},
			pdfsubject={Projeto de Pesquisa Crime e Blitz},
			pdfcreator={Marcelino Guerra},
			pdfkeywords={abnt}{latex}{abntex}{abntex2}{atigo científico}, 
			colorlinks=true,       		
			linkcolor=blue,          	
			citecolor=blue,        		
			filecolor=magenta,      		
			urlcolor=blue,
			bookmarksdepth=4
		}
		
\begin{document}
			
			\frenchspacing 
			
			\maketitle
			\begin{abstract}
				
				We evaluated one of the most common policing strategies in Brazil: the allocation of police blitzes. This place-based focused deterrence intervention has well-defined assignments, and 3,423 interventions were precisely recorded in Fortaleza-CE, Brazil, between 2012 and 2013. Our analysis takes advantage of the high spatiotemporal daily data resolution coming from an unprecedented longitudinal micro-Big Data (GPS and PING records) to make comparisons of small intervention areas, while controlling for common daily trends, deterrence (spatial and temporal), and diffusion; to show that an average police crackdown causes a 35\% decrease in violent crime occurrences. There are diminishing returns of public safety to hours spent by police in a single area, corroborating what police officers know well from their own experience and discretionary behavior. Although crime increases by 6\% immediately after the end of a blitz, we observe lasting deterrent effects (diffusion) after 2-3 days. The residual deterrence cancels the relocation of the crime, and the intervention does not generate significant temporal displacement. In addition, we do not find spatial displacement from crime in blocks up to 1.5 km from a blitz. This type of micropolicing tactics generates deterrence by being highly visible in a street segment for a short period and intermittently quasirandom in space-time, which produces uncertainty that might be crucial in minimizing the temporal and spatial displacement of crime. Of public policy interest, we show that the allocation of blitzes passes in an initial cost-benefit analysis.\\
				
				%
			\end{abstract}
			
			\section{Introduction and Background}\label{SECTION:Introduction}
			In the last three decades, Latin America has been described as an unsafe and violent region. This is the case, despite all the socioeconomic progress experienced by the region, see \citeonline{Chioda2017-wb}. Throughout this time, cities in the region have consistently ranked among the 50 most violent places in the world, and by 2021, a total of 38 cities belong to the region\footnote{Indeed, in 2013, of the top 50 most violent cities in the world, 42 were in the region, including the top 16.}. Not less than nine cities on this list are in Brazil, accounting for almost 20 percent of the 2021 ranking, \citeonline{Lista50}. 
			
			More specifically, Fortaleza (a large coastal city, the capital of the state of Ceará in northeastern Brazil) has consistently ranked on that list. The city is considered the ``crown jewel'' of transnational traffic routes to North Africa and Europe; and, as a consequence, Brazil’s two main gangs, the First Capital Command (Primeiro Comando da Capital – PCC) and the Red Command (Comando Vermelho - CV) have spent considerable efforts there in turf wars that profoundly transformed the landscapes of social, homicidal, and violence in the city. In fact, empirical studies have shown the role that gangs play in shaping crime rates in large cities of Latin America by promoting the diffusion of violent crimes, especially through the formation of social networks with the objective of killing members of rival gangs or controlling local drug markets; see \citeonline{De_Oliveira2019-vl}, see \citeonline{SullivanGANGS2022} and \citeonline{BunkerSulivan2022} and the references therein. 
			
			Not surprisingly, there is considerable interest among public policy makers in acquiring knowledge and expertise in the prevention of urban crime and violence. The local government police authorities are continually seeking to identify innovative interventions that can be implemented, managed, and assessed on a micro-urban scale.
			
			Since there is much more evidence that crime is
			responsive to police compared to alternatives such as attractive legitimate labor market opportunities or the severity of criminal sanction, see \citeonline{chalfin2017criminal}; it is understandable that policing are gaining prominence these days. In fact, \citeonline{Weisburd2021-eu} points out that, today, innovative crime prevention programs tend to focus on high doses of deterrence in small areas or over short time periods, i.e., hot spot policing, pulling-levers policing, police crackdowns, community interventions via neighborhood policing or broken-windows policing.
			
			Such type of micro-policing focused deterrence can reduce violence in urban hot spots and is very welcomed as local governments face diminishing police resources resulting from tight city budgets and scarce federal money. However, these authorities are aware of the need to adopt only innovations whose impacts have been validated by rigorous research and empirical evidence brought about by scientific traditions like crime economics, especially those that are thoroughly framed on the preventive role of deterrence.
			
			Most of the crime economics and criminology research focuses on estimating the deterrent effect. Following \citeonline{becker1968crime} seminal work, which models criminal behavior as a cost-benefit calculus, the efforts of economists and criminologists were devoted to credibly establishing causality between crime and punishment, as well as incentives (positive or negative). The relevance of knowing the impact of treatments and interventions in crime for criminologists is epitomized by the voluminous and influential academic literature on what has been termed the `What Works' movement, see \citeonline{ Weisburd2017-nx}.
			
			From a predominant narrative in crime prevention and rehabilitation that nothing works, nowadays Law \& Order institutions have more than 40 years of theoretical, methodological and evaluative developments on `What Works' to combat crime and disorder, broadly classified into the following tags (\citeonline{ Weisburd2017-nx}): developmental and social prevention, community interventions, situational prevention; policing, sentencing, correctional interventions and drug treatment and interventions.
			
			Although policing encompasses any police programs, training, interventions, or activities (for example, patrolling beats, police crackdown, police station allocation, police-monitored cameras, police body-worn cameras, etc.) that sought to impact crime and other police outcomes; our analysis here is focused on a specific type of policing strategy, commonly referred to as police crackdowns.
			
			Since policing often reacts to higher levels of crime, creative research designs have been used to avoid that endogeneity\footnote{As to economists contributions, see \citeonline{levitt1997using}, \citeonline{levitt2002using}, \citeonline{di2004police}, \citeonline{klick2005using}, \citeonline{evans2007cops}, \citeonline{draca2011panic}, and \citeonline{mello2019more}, among others. Regarding the voluminous and seminal contributions of criminologists, see \citeonline{Kelling1974-cq}, \citeonline{Sherman1984-mm}, \citeonline{Sherman1995-ky}, \citeonline{Taylor2011-fq}, \citeonline{Braga2001-bm}, \citeonline{piza2014saturation}, \citeonline{Petersen2023-fb}, \citeonline{Weisburd2023-wv}, among many others.} and show the causal effect of police on crime. However, these studies do not have a definitive opinion on what more police would exactly do on the streets to stop unlawful behavior beyond making arrests (see  \citeonline{nagin2013deterrence}), nor on how such police interventions impact crime both spatially and temporally; supporting a rather pessimistic consensus about the likely impacts epitomized by \citeonline{Guerette2009-rh}'s say that ``despite emerging evidence to the contrary, the prevailing sentiment seems to be that crime displacement is inevitable'. 
			
			Relocation of crime across space and time due to changes in patrolling (displacement) is compatible with the rational approach toward deviance but has received less attention from the field. For example, offenders facing a higher probability of apprehension caused by police presence in one area may relocate their criminal activities to blocks farther away to avoid arrests or wait until cops are gone to commit a crime.
			
			According to \citeonline{Guerette2009-rh}, displacement is the relocation of a crime from one place, time, target, offense, tactic, or offender to another (say, five dimensions of change) as a result of some crime prevention initiative. Implicitly in that definition, it is a subtle challenge to crime interventions, i.e. although a specific crime intervention might look like effective (by decreasing crime statistics under a specifically defined unit of analysis), such change is elusive, as what really could have happened was just a change on one of those dimensions. In criminology, spatial displacement is the form most commonly recognized and analyzed followed (very behind) by temporal displacement, see \citeonline{Guerette2009-rh}, \citeonline{Bowers2011-wm}, \citeonline{Telep2014-jr}, and \citeonline{Hatten2022-xo}. 
			
			Diffusion, another key criminological definition, occurs when crime reductions (or other improvements) are achieved in areas that are close to crime prevention interventions, even though those areas were not actually targeted by the intervention itself; see \citeonline{ClarkeWeisburd1994} and \citeonline{ Telep2014-jr}.  Diffusion has other names such as bonus effect, halo effect, free rider effect, multiplier effect, and spillover effect.
			
			Whenever the researcher ignores temporal and spatial displacement, the direct effects of interventions will undoubtedly overestimate total effects. In the condition that crime entirely moves around the corner or clock, place-based policing assignments are doomed to failure. In terms of diffusion, ignoring it has an opposite impact on the measurement of total effects.
			
			Technically, the above-mentioned issue is called spillover effects by economists, i.e., when an intervention, program, or treatment targeted at a specific population (or unit of treatment, UT) affects other populations (or units of treatment). Usually, in that context, well-defined and non-intersecting geographic areas are chosen as UTs. Then, some UTs receive treatment while others do not (control areas); and the objective is to assess the impact of that treatment on crime ocurring in UTs and possibly in nontreated areas.
			
			Spillover effects have been recently explored by many different disciplines\footnote{In economics, see \citeonline{Baird2015-ld}, \citeonline{Angelucci2016-oi}, \citeonline{Fletcher2017-ba}, \citeonline{Muir2020-ux} and \citeonline{Mendoza-Jimenez2024-dr}, in regional science, see \citeonline{halleck2015slx}; and in criminology, see \citeonline{Munyo2020-bb}, and \citeonline{Vilalta2023-ah}.}, and its empirical content raised key methodological issues. The impossibility of interference between treatment units has traditionally been assumed by randomized experiments, since standard experimental designs cannot identify and measure spillovers. However, if spillover exists, a new empirical approach is needed if one wants to identify and measure impacts correctly, see \citeonline{halleck2015slx}, \citeonline{Angelucci2016-oi}, \citeonline{Aronow2021-mr} and \citeonline{DiTraglia2023-uk}.
			
			As a specific strategy of policing, police crackdowns consist of a geographically focused and offense-specific proactive policing assignment and are another popular tactic\footnote{For an evaluation of different tactics and innovation in police practices, see \citeonline{weisburd2004can} and \citeonline{chalfin2017criminal}.} to enforce the law. \citeonline{sherman1990police} evaluates eighteen case studies of crackdowns on prostitution, drug trafficking, and drunk driving, among others. He observes that most papers find direct impacts on crime (successful initial deterrence), and some also detect residual deterrence: even after the police operation, crime levels remained lower. In addition, crime displacement (or spillover effects) was found in four studies related to drug trafficking crackdowns.   
			
			That said, this paper aims to evaluate one of the most common policing strategies in Brazil: the allocation of blitzes, our version of the police crackdown. Officers block street segments for five to six hours, stopping and searching around 100 vehicles, commonly distributing fines for traffic transgressions. In essence, blitzes are police crackdowns organized by the State Highway Police (Polícia Rodoviária Estadual) to inhibit and punish driving violations, such as driving under the influence of alcohol or illegal substances, and invalid licenses and vehicle registration. As a branch of the Military Police, they also stop and search vehicles to make arrests and apprehend drugs and weapons.
			
			The command aims to surprise drivers by assembling blitzes in different street segments at different times and wants to create a reputation for fighting traffic violations relentlessly. In addition, the police focus on specific segments of the streets, but vary the intensity and timing of the treatment to establish the fear of presence at any time.  Our study area is Fortaleza. Between 2012 and 2013, we identified 3,423 unexpected blitzes\footnote{This represents more than 19,000 hours of diligent police work on the streets over two years.} and 68,243 violent crimes across street segments of Fortaleza, a large Brazilian city. This kind of policing assignment targets roadway safety and does not react to crime waves, alleviating concerns about simultaneity.
			
			These blitz operations came as a sort of coordinated reaction to a considerable and threatening increase in violent crime throughout the city. As mentioned in \citeonline{De_Oliveira2019-vl}, at that time, the trends in violent crimes (murders, bodily injury, or theft that caused the victim's death) in Fortaleza were steadily increasing. Fortaleza responded, on average, to 43. 8\% of all violent crimes in the state of Ceará and 3.0\% of all violent crimes in Brazil between 2009 and 2015. Violent crimes increased substantially during the period, especially in the state capital. The number of occurrences doubled between 2009 and 2013, jumping from 977 to 1993 in Fortaleza. The violent crime rate per 100,000 inhabitants increased from 39.0 to 78.1 between 2009 and 2013, a variation of 100\%. In 2015, the rate dropped to 63.3 per 100,000 inhabitants, but it was still almost 2.5 times higher than the rate for the entire country.
			
			We take advantage of the high spatiotemporal resolution of the data (both for the location of blitzes and the spatial distribution of crime) to make comparisons of small intervention areas in the same period of day and day of the week while controlling for common daily trends, for displacement (spatial and temporal) and for diffusion effects; and show that an average police crackdown causes a 35\% decrease in violent crime occurrences. As expected, there are diminishing returns of public safety to the hours spent by police in a single area. Although crime increases by 6\% immediately after the end of a blitz, we observe lasting deterrent effects after the next 2 and 3 days, i.e. diffusion effects or residual deterrence. In general, residual deterrence cancels out crime relocation, and the intervention does not generate significant temporal displacement. In addition, we do not find spatial relocation of violence to blocks within 1.5 km of a blitz. 
			
			In two years, the blitzes generated US\$ 2.3 million in public safety improvement. In fact, US\$ 3.2 million operational costs were surpassed by US\$ 7.4 million in revenues related to traffic tickets. Even without considering the potential reduction in traffic fatalities caused by the policy, blitzes certainly pass in the cost-benefit analysis. Finally, highly visible, short-term, and intermittent police crackdowns appear to minimize crime displacement while reducing violent crimes.
			
			
			The novelty of our analysis comprises at least 4 dimensions: 1) assessing the total impact - direct and indirect (spillover effects) - of a well-defined and large-scale (entire city) police operation; 2) using as empirical benchmark a combination of very disaggregated and precisely measured GPS crime data coupled with PINGs related to police blitzes spatial and time dynamics; 3) estimating the presence of spatial displacement by incorporating inverse distance weight matrices that identify suitable catchment areas that include nearby, intermediate, and distal crime movements; and 4) measuring temporal displacement or lasting deterrent effects of place-based interventions by exploiting the high frequency of the data using multiple temporal lags.
			
			Besides this Introduction, there are five additional Sections. Section \ref{SECTION:LitReview} reviews the recent literature on crime displacement made by both criminologists and economists with special emphasis on those attempts that applied rigorous quantitative methodology to carefully collected microdata sets. Section \ref{SECTION:DataContext} provides details on the general socioeconomic and criminological profile of the city of Fortaleza, as well the way blitzes are conceptualized, spatially and time allocated, the difference between the two main types of blitzes (``fixed'' \textit{versus} ``mobile''), and the process of data gathering and sample construction for analysis.
			
			In Section \ref{SECTION:ResearchDesign}, we carefully describe the identification and estimation strategy employed. As to our identification step,  note that police command aims to surprise drivers (including potential offenders) in space-time and give them the impression that they could be subject to stop and search at any time, i.e. drivers perceived a quasirandom allocation of treatment. Hence, by design, this police assignment does not react to previous crime waves, alleviating concerns about simultaneity. In addition, the high spatial resolution and high time frequency create additional opportunities to isolate the direct and indirect causal effects of policing on crime. To estimate policing spillovers, we build a 2,562 by 2,562 spatial weight matrices based on the inverse distance of cells' centroids to each other and apply different specifications of Poisson regressions that include time lags. Section \ref{SECTION:Results} shows all estimated results and describes our attempts to unravel the two main mechanisms through which we believe these local police interventions could affect crime occurrences: 1) deterrence and 2) a type of incapacitation effect (catching drivers who have targeted the area surrounding the blitz for criminal activities). Also, it presents two additional empirical exercises, i.e. a robustness check to the size of spatial grid cells and a cost-benefit analysis. Finally, Section \ref{SECTION:FinalRemarks} concludes and makes some suggestions for future improvements in our analysis.
			
			\section{Policing, Crime Displacement and Diffusion}\label{SECTION:LitReview}
			
			\citeonline{johnson2014crime} review the criminological literature on crime displacement and point out that, most likely, crime does not simply move around the corner, since the empirical evidence finds that diffusion of crime control benefits occurs at least as frequently as crime displacement. However, research on this matter usually deals with small-scale interventions, restricted only to a subset of a city, and only considers the immediate surrounding areas where a police intervention occurs. The choice of the catchment area is crucial because the regions most likely to accommodate crime relocation may be farther away from adjacent blocks. In addition, only a few studies explored the possibility of temporal displacement.
			
			In a recent review of the literature, \citeonline{braga2019hot} show that the vast majority of studies find that focusing resources on high-crime places (``hot spots'') generates relevant crime control gains without displacing illicit activities to nearby areas. However, only a portion of these studies attempt to measure spatial diffusion (or spatial spillover effects). The focus on ``hot spots'' allocation of police resources is not new. In fact, the pioneering analysis of \citeonline{sherman1989hot} revealed that crimes disproportionately focus on a few streets, the ``hot spots'', suggesting that targeted policing assignments could be a cost-effective way to address the problem. With the help of new mapping technologies, hot spot policing has become commonplace in the past 30 years. 
			
			Perhaps one of the most surprising results comes from the carefully performed analysis of place-based interventions by \citeonline{blattman2021place}. The authors evaluate a large-scale experiment in Bogot{\'a}, Colombia. For eight months, 1,919 street segments with moderate to high crime levels were assigned more police or more state presence\footnote{More state presence in their context means, for example, street lights maintenance, graffiti cleanup, and trash collection.}, both or none. Taking into account spatial spillovers within 250 meters of treated hot spots, the study estimates that about 872 crimes are displaced to nearby streets, which is not new in the literaure\footnote{For instance, \citeonline{piza2014saturation} estimate of both temporal and spatial displacement of robbery from Operation Impact, a foot-patrol initiative in Newark, NJ. In contrast to \citeonline{blattman2021place}, the authors find direct effects of increased police patrols on aggravated assault, shootings, and overall violence}. However, the fact that 84\% increased police patrolling had negligible direct effects on crime occurrences in a small intervention area is unforeseen. 
			
			\citeonline{Smith2024-uv} evaluates the impact of a hot-spot policing strategy in Dallas, Texas. Their strategy was implemented in 2021 by the Dallas Police Department (DPD) as part of a city-wide strategy to reduce violent crime. While most hot spots experiments are relatively short in duration, the Dallas Crime Plan lasts over an extended period of 3–5 years. In addition, another interesting feature is the broadening of the intervention concept of hot spots to include both efforts to impact violent places and violent people. They used a difference-in-differences approach (which allows consideration of temporal and spatial patterns of crime) and found evidence that violent crime fell, on average, by 11\% in the hot spots targeted during the first year of the intervention without evidence of spatial crime displacement. Last, targeted hot spots to city-wide violent crime decreased significantly over the course of the year, corroborating the findings of diminishing returns of public safety to the intervention.
			
			In addition, there is a recent but growing literature (see \citeonline{Vidal_undated-ip} and \citeonline{Weisburd2021-eu}) that attempts to provide an estimate of the social returns (impact) of policing on crime levels that takes advantage of large amounts of fine-grained and precisely georeferenced data about criminal activities tracked both by traditional GPS techniques and by information coming from thousands (sometimes millions) of PING\footnote{A PING (Packet Internet or Inter-Network Groper) is a basic Internet program that allows a user to test and verify if a particular destination IP address exists and can accept requests in computer network administration. This technology is key to tracking the trajectories of moving objects and vehicles. It is commonly used by police departments worldwide.} information related to police space-time dynamics. Indeed, such type of \textit{Big Data} use is a clear trend in both economic \& econometric (\citeonline{Varian2014-au}, and \citeonline{LIN2024254}) and criminal (\citeonline{Su2023-tz}, and \citeonline{Lee2024-js}) academic applied circles to improve exploratory data analysis, forecast, causal modeling, and simulation.
			
			\citeonline{Vidal_undated-ip} exploits a set-up that aimed to increase police presence in more than 6,000 well-defined areas in the city of Essex, United Kingdom. In October 2013, Essex Police introduced a new operation that, over a period of 19 months, targeted these areas. Every week a different set of areas was chosen, with each area receiving an average of ten additional minutes of police presence per day. Together, these areas represented the population of locations where crime (specifically, private dwelling burglary) occurred in Essex during this period. Using data transmitted by GPS devices worn by police officers, they documented exogenous and discontinuous changes in patrol intensity. They do not find that these increases in patrolling were accompanied by corresponding decreases in crime. 
			
			\citeonline{Weisburd2021-eu} estimates the impact of police presence on crime using a unique database that tracks the exact location of Dallas Police Department patrol cars throughout 2009, in the United States. Since the 1930s, police patrol in US cities has been dominated by the rapid response system: a policing strategy that allocates patrol cars driving around in police beats ready to respond rapidly to an emergency call; when they are not responding to such calls, they spend their time in what has been termed random preventative patrol, showing their presence in the beats to deter offending. She examines the impact of this rapid response strategy on deterrence. To address the endogeneity concern that the location of the officer is often driven by crime, the author exploits police responses to calls outside of their assigned coverage beat. She finds that a 10 \% decrease in the presence of police at that location results in a 7 \% increase in crime.
			
			To our knowledge, \citeonline{Vidal_undated-ip} and \citeonline{Weisburd2021-eu}  are the only two attempts so far to look at the geographic distribution of police officers throughout an entire area of a city using precise GPS-level data on police location and assess its impact on crime. In fact, our current analysis of blitzes shares an analogous geographic scope and goal. However, ours is the only one applied to a very high crime area represented by a large urban center (2.5 million people living in a 121 square miles area) in a developing world city, Fortaleza (state of Ceará, Brazil), which carefully models spatial and temporal displacement, as well as diffusion effects.

			\section{Data and Context}\label{SECTION:DataContext}
			
			\subsection{Fortaleza-CE, Brazil}
			
\begin{figure}[!h]
				\centering 
				\includegraphics[scale=0.5]{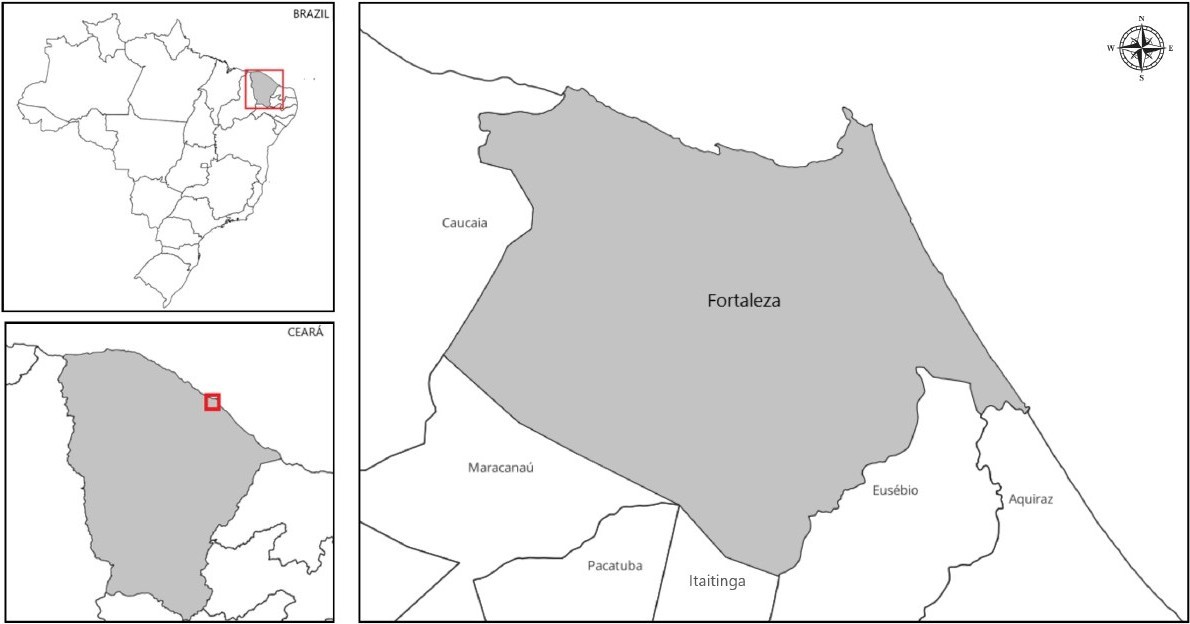}
				\caption{The City of Fortaleza-CE on the Map}
				\label{FIGURE:MapFortaleza}
			\end{figure}

			Fortaleza is the capital and most populous city of the state of Ceará, located in the Northeast region of Brazil, 2,285 km from Brasília, Brazil's federal capital, see Figure \ref{FIGURE:MapFortaleza}. It is located on the Atlantic coast, has an area of 313,82 km² and 2.5 million inhabitants in 2022, making it the largest city in the North and Northeast regions in terms of population and the fourth largest in Brazil. The city has the highest population density among the country's capitals, with 8,390.76 inhabitants per square kilometer.
			
			\noindent In terms of GDP per capita and Human Development Index (HDI), Fortaleza is listed at 22$^{nd}$ and 19$^{th}$ among Brazilian cities, respectively, and has a low average per capita income; see Table \ref{TABLE:AverageIncome}. In fact, at the bottom of Table \ref{TABLE:AverageIncome} we see that the ten cities with the lowest average monthly household income per capita in 2023 belong to the North or Northeast regions of the country, and Fortaleza ranks very low at 20$^{th}$.
			
			\begin{table}[!h]
				\footnotesize
				\centering
				\caption{Average per capita monthly household income for Brazilian States Capital Cities, 2023}
				\label{TABLE:AverageIncome}
				\begin{tabular}{llcc}
					\tabularnewline \midrule \midrule
					~ & City & State & Income (US\$)\\
					\hline
					1 & Florianópolis & SC & 819.85 \\ 
					2 & Vitória & ES & 740.58 \\
					3 & Porto Alegre & RS & 709.12 \\
					4 & Rio de Janeiro & RJ & 677.21 \\ 
					5 & São Paulo & SP & 653.26 \\ 
					6 & Brasília & DF & 642.97 \\ 
					7 & Curitiba & PR & 636.32 \\ 
					8 & Belo Horizonte & MG & 623.32 \\ 
					9 & Goiânia & GO & 569.23 \\ 
					10 & João Pessoa & PB & 543.02 \\ 
					11 & Cuiabá & MT & 508.42 \\
					12 & Palmas & TO & 502.56 \\ 
					13 & Natal & RN & 463.03 \\ 
					14 & Campo Grande & MS & 456.94 \\ 
					15 & Belém & PA & 452.40 \\ 
					16 & Teresina & PI & 436.30 \\ 
					17 & Macapá & AP & 377.72 \\ 
					18 & Aracaju & SE & 377.07 \\ 
					19 & Salvador & BA & 364.74 \\ 
					20 & Fortaleza & CE & 354.29 \\ 
					21 & Recife & PE & 324.20 \\ 
					22 & Boa Vista & RR & 324.03 \\ 
					23 & Maceió & AL & 319.47 \\ 
					24 & São Luís & MA & 313.92 \\ 
					25 & Porto Velho & RO & 305.57 \\ 
					26 & Manaus & AM & 299.93 \\ 
					27 & Rio Branco & AC & 263.75 \\
					\midrule \midrule
					\multicolumn{4}{l}{\footnotesize{Source: \citeonline{SocialIndic}.}} \\
					\multicolumn{4}{l}{\footnotesize{Exchange rate 1.0 US\$ = 5.0 R\$, annual average}} \\
					\multicolumn{4}{l}{\footnotesize{rate in 2023 calculated by Banco Central (BACEN),}}
					\\
					\multicolumn{4}{l}{\footnotesize{Brazilian monetary authority.}}
				\end{tabular}
			\end{table}
			
			\noindent Income inequality remains high in Fortaleza, with a Gini coefficient of 0.61. According to data from 2022, around 22.9\% of Fortaleza's population lives in \textit{favelas}\footnote{A \textit{Favela} is a slum or shantytown located within or on the outskirts of the largest cities of the country characterized above all by extreme porverty. In these places there is a widespread lack of infrastructure, rudimentary water and sewage capacity, unsanitary conditions, poor nutrition, pollution, disease, and infant mortality rates are high.}, making it the seventh most \textit{favela}-ridden capital in the country. Frequent droughts and the resulting rural exodus from the interior of the Ceará state aggravate the problem of \textit{favelas}. Despite that, Fortaleza is one of the main tourist destinations in the country. According to the Ministry of Tourism, in 2012, it was the second most desired destination in Brazil and the fourth that received the most visitors\footnote{Although this activity generates considerable income for the city, there is a dark side to it: the combination of social vulnerability and absence of public authorities has consolidated Fortaleza as a destination for (national and international) sex tourism and child prostitution, see \citeonline{Aquino2015-ln}.}.
			
			The city is commonly ranked as one of the most dangerous places in the world. Homicide rates were around 68 per 100,000 people in 2012, much higher than Cali-Colombia, St Louis-USA, New Orleans-USA, and Baltimore-USA, and twice as large as Brazil's 25-30 murder rates per 100,000 people. Most violent crimes are related to robberies, which reached more than 130 daily occurrences in 2012. More recently, figures continue to support this trend; see Table \ref{TABLE:HomicideRates}. 
			
			\begin{table}[!h]
				\footnotesize
				\centering
				\caption{Homicide Rate (per 100,000) for Brazilian States Capital Cities, 2022}
				\label{TABLE:HomicideRates}
				\begin{tabular}{llcccc}
					\tabularnewline \midrule \midrule
					~ & City & State & Region & Homicide Rates 2012  & Homicide Rates 2022  \\ \hline 
					1 & Salvador & BA & NE & 61.6 & 66.4 \\
					2 & Macapá & AP & N & 36.0 & 55.8 \\ 
					3 & Manaus & AM & N & 54.0 & 55.7 \\ 
					4 & Porto Velho & RO & N & 40.1 & 47.6 \\ 
					5 & Fortaleza & CE & NE & 71.5 & 45.3 \\ 
					6 & Recife & PE & NE & 40.2 & 44.7 \\ 
					7 & Aracaju & SE & NE & 41.9 & 41.8 \\ 
					8 & Maceió & AL & NE & 78.3 & 41.5 \\ 
					9 & Teresina & PI & NE & 35.7 & 40.4 \\ 
					10 & Boa Vista & RR & N & 27.0 & 39.2 \\ 
					11 & Natal & RN & NE & 48.9 & 36.9 \\ 
					12 & Palmas & TO & N & 18.5 & 32.0 \\ 
					13 & Porto Alegre & RS & S & 36.9 & 29.0 \\ 
					14 & Vitória & ES & SE & 38.2 & 28.5 \\ 
					15 & São Luís & MA & NE & 52.7 & 27.2 \\ 
					16 & Belém & PA & N & 54.1 & 26.5 \\ 
					17 & Rio Branco & AC & N & 27.8 & 25.8 \\ 
					18 & João Pessoa & PB & NE & 65.1 & 23.5 \\ 
					19 & Rio de Janeiro & RJ & SE & 20.6 & 21.3 \\ 
					20 & Curitiba & PR & S & 32.7 & 21.0 \\ 
					21 & Campo Grande & MS & CO & 21.7 & 19.8 \\ 
					22 & Belo Horizonte & MG & SE & 35.0 & 17.6 \\ 
					23 & Goiânia & GO & CO & 45.5 & 16.1 \\ 
					24 & São Paulo & SP & SE & 16.3 & 15.4 \\ 
					25 & Cuiabá & MT & CO & 42.2 & 15.2 \\ 
					26 & Brasília & DF & CO & 35.1 & 13.0 \\ 
					27 & Florianópolis & SC & S & 14.0 & 8.9 \\
					\midrule \midrule
					\multicolumn{6}{l}{\footnotesize{Source: IPEA and \citeonline{AtlasViolencia2024}.
					}} \\
					\multicolumn{6}{l}{\footnotesize{Brazil has 27 sates and a Federal District (Brasília) grouped into}} \\
					\multicolumn{6}{l}{\footnotesize{five different geographical regions, i.e. Northeast (NE), North (N),}} \\
					\multicolumn{6}{l}{\footnotesize{Southeast (SE), Center-West (CO) and South (S).}}
				\end{tabular}
			\end{table}
			
			\noindent Compared to other major Brazilian cities, Fortaleza is usually the top of the list of violent crimes, even among those in the Northeast region. In the last 30 years, the cities of the north and northeast have exceeded the levels of violence in the southeast and are currently the most dangerous areas to live in Brazil, with homicide rates of around 40 per 100,000 people; see Table \ref{TABLE:HomicideRates}.
			
			A strongly supported explanation for this homicidal prominence of cities in the Northeast and North stems from a profound change in the criminal and homicidal dynamics of these regions that began in the early 2000s and continues today. There has been a significant strengthening and consequent intensification of clashes between the two major organized drug trafficking groups (gangs) in Brazil, that is, the First Command of the Capital (Primeiro Comando da Capital - PCC) and the Red Command (Comando Vermelho - CV)\footnote{Both PCC and CV are resilient transnational organized crime groups from Brazil with paramilitary structures and budgets among the largest in the world. In fact, President Biden signed an Executive Order (EO 14059 on December 15, 2021; see https://ofac.treasury.gov/recent-actions/20211215) to modernize the US Department of Treasury sanctions authorities used to combat illicit drug trade, and the Department of Treasury designed 25 actors, listing the Brazilian PCC. } to control their entry routes in the northern region and their exit routes in the northeast region. This war pitted the PCC and its smaller local allies against the CV. Practically all states in the northern region have had their homicide dynamics profoundly affected as a result of the struggle to conquer the traffic routes. Amazonas, Pará, Amapá, Acre and Roraima were the hardest hit. Downstream of the route, at those points where trafficking leaves the national territory for North Africa and Europe, the impacts were also significant, placing port cities in the Northeast such as Fortaleza, Natal, and Salvador in a situation of calamity in terms of public security that continues to this day; see \citeonline{Berg2021-mh} and \citeonline{Sullivan2022-hb}.
			
			Regarding policing levels, the Polícia Militar do Ceará - PMCE (Military Police of Ceará) staff had around 14.900 officers in 2012, one policeman per 565 people. This ratio is more favorable in the capital since around 50\% of the staff is allocated in Fortaleza, but it is still considered low compared to cities of similar size. It is important to emphasize that the Military Police, responsible for preventive policing and first response to calls, have different branches that employ distinct strategies such as community policing, regular patrolling, and traffic law enforcement, among others. 
			
			\subsection{The Allocation and Operation of Blitzes}
			
			In many Brazilian cities, blitzes are a common policing strategy to combat deviant behavior. In general, police officers are standing in a linear row and block streets using traffic cones. Using visual and audible warnings, police choose several vehicles in targeted segments for checks and inspections. Truly, they are a specific type of ``police crackdown'' (see, \citeonline{sherman1990police}), where police officers and a group of policemen and other accessory public employees (transit authority, county civil guard, public health professional, and others) are moved as large groups in focus areas, representing a sudden and dramatic increase in police officer presence and sanctions for all offenses in (random) specific places. 
			
			\begin{figure}[!h]
				\centering 
				\subfloat[Blitz with policemen only]{{\includegraphics[width=6.5cm, height=4cm]{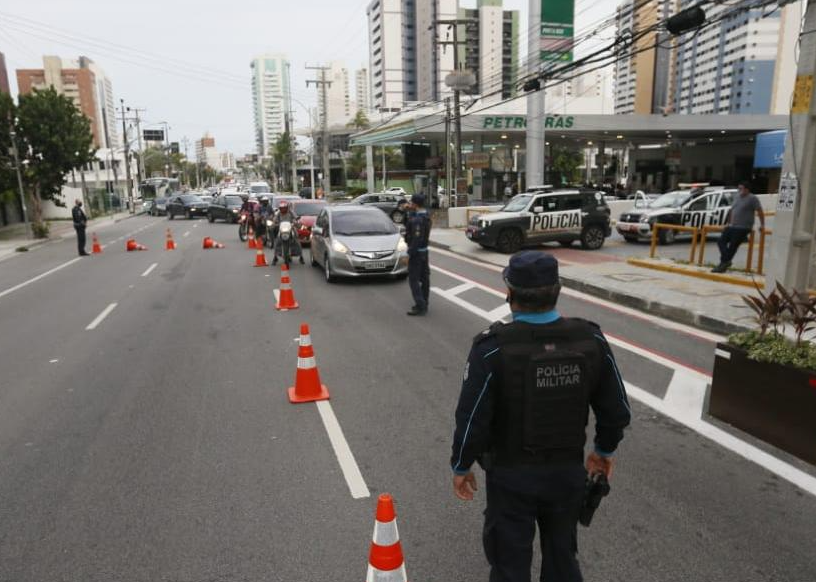}}}
				\quad
				\subfloat[Blitz with policemen and transit authorities]{{\includegraphics[width=6.5cm, height=4cm]{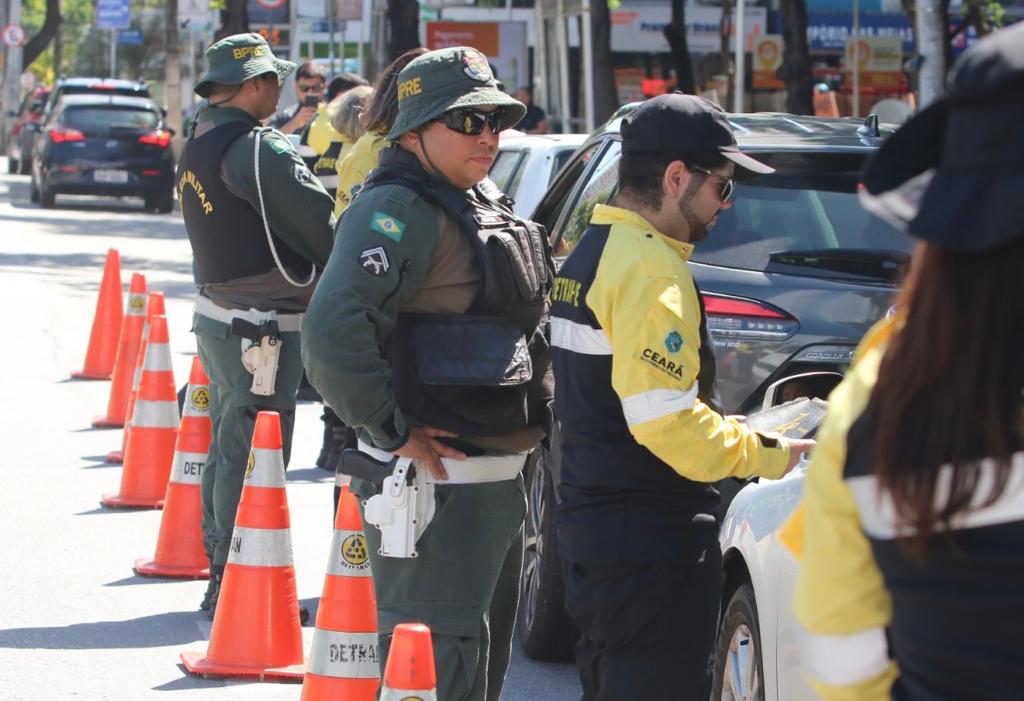}}}
				\\
				\subfloat["Stop-and-Frisk" Blitz]{{\includegraphics[width=6.5cm, height=4cm]{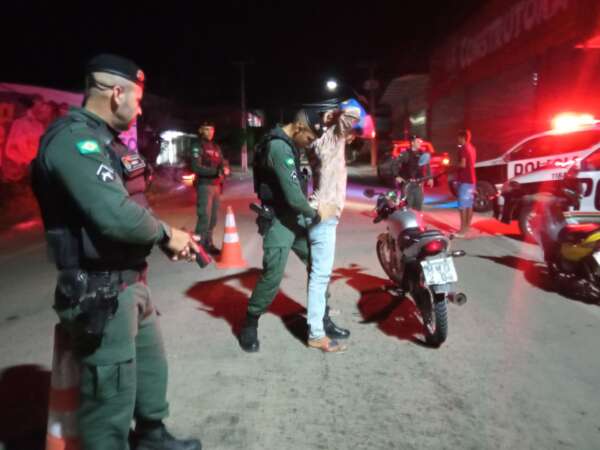}}}
				\quad
				\subfloat[Blitz during 2024 Brazilian Carnival]{{\includegraphics[width=6.5cm, height=4cm]{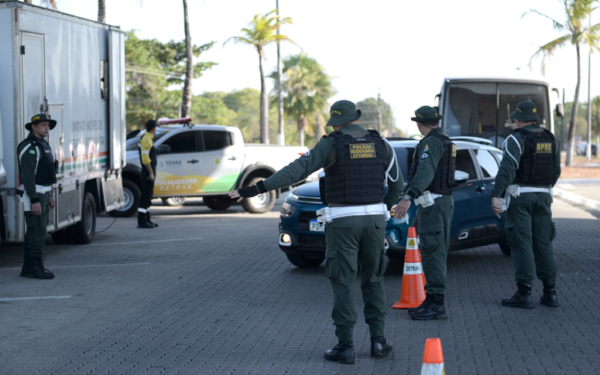}}}
				\caption{Examples of Blitzes}
				\label{FIGURE:RealBlitzes}
			\end{figure}
			
			\noindent Brazilian blitzes protocols encourage officers to use a broader range of tactics to address many crime and illicit related problems, exercising full discretion and initiative, including when finishing the operation. In addition to taking more enforcement actions, police officers can also be encouraged to perform specific actions such as arresting offenders; issuing citations; conducting field interviews; issuing written or verbal warnings; conducting highly visible patrols; conducting traffic stops (indeed, this is exactly the first action performed; after that stop, officers initiate a full set of legal procedures on the occupants of the vehicle and the vehicle itself); inspecting driver’s licenses; searching vehicles and interviewing drivers at roadblocks or checkpoints; and checking for guns and drug possession.
			
			Although there is some discretion on which vehicle to stop, commanders give general directives such as stopping motorcycles, especially when there is also a passenger. They aim to reduce murders, since a shooter on the back of a motorcycle performs most of the killings. This stop and search procedure can size vehicles, weapons and drugs, and cops frequently distribute fines for driving offenses\footnote{Most drivers stopped in a blitz are subject to Blood Alcohol Content (BAC) tests, and car inspections might find traffic violations such as broken tail lights, expired driver license, among others.}. In addition, all policemen on duty at any blitz are obligated to act on any flagrant crime or disorder and to attend any emergency around the premises of the operation. Figure \ref{FIGURE:RealBlitzes} shows some pictures of real blitzes.
			
			Blitzes can produce both deterrence and incapacitation, see \citeonline{nagin2013deterrence} and \citeonline{chalfin2017criminal}. In terms of deterrence, by being randomly allocated, acting visible, conducting traffic stops, inspecting vehicles and people, and being obligated to act on any incivility around the premises of the operation; blitzes change criminals (actual or potential) behavior as their probability of catchment increases. As to incapacitation, if during or after the full set of legal procedures conducted by blitzes policemen upon the occupants of the vehicle and the vehicle itself, any illegality is characterized (DUI, illegal possession of fire arms, vehicle out-of-transit regulations, reckless driving, suspect behavior, open arrest warrant, and others); all law-breaking occupants are immediately detained and sent to a nearby police station for further legal procedures. 
			
			The State Secretary of Public Security provided information about the universe of blitzes in Ceará state during 2012 and 2013; in fact, José Raimundo Carvalho worked closely with the Bureau of Operations to collect and compile as much information as possible regarding the universe of Police crackdowns in Cear{\'a} State between 2012 and 2013. The first intervention occurred in January $9^{th}$, 2012, and the last in December $15^{th}$, 2013. Most records have the exact times the blitz started and ended, full street addresses, and numbers of officers and police vehicles allocated. Information about cars and motorcycles stopped and vehicles, weapons, and drugs seized is also available. A total of 3,423 interventions could be geolocated\footnote{In total, there were 3,656 interventions in Fortaleza from 2012 to 2013. Some do not have complete addresses, and some lack information about start and end times. The ones missing crucial information are dropped from the sample.} within Fortaleza's boundaries, adding to 19,085 police work, or 27 hours of policing per day. 
			
\begin{figure}[!h]
				\centering 
				\includegraphics[width=12cm]{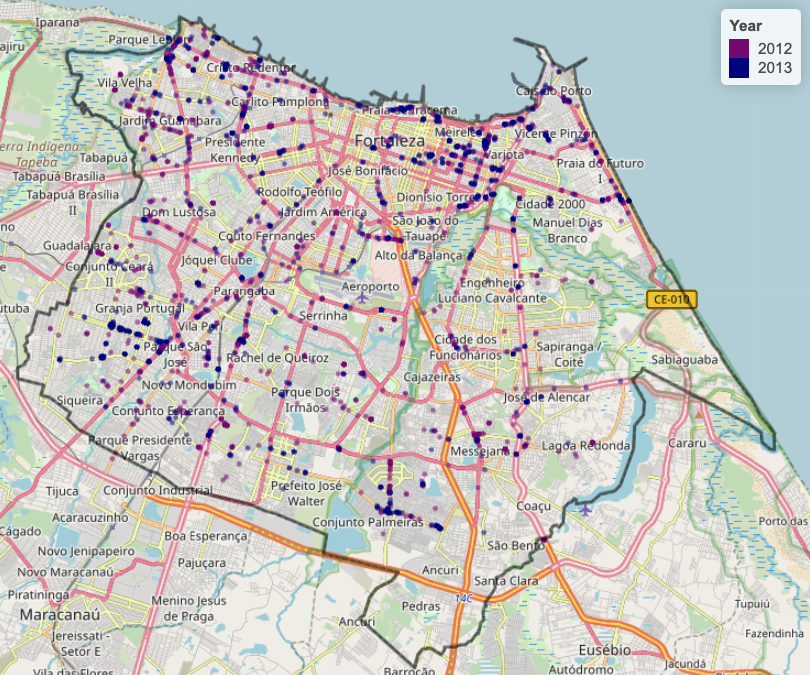}
				\caption{Spatial Distribution of Blitzes in Fortaleza-CE, 2012-2013}
				\label{FIGURE:SpatialDistBlitzes}
			\end{figure}

			This kind of policing strategy has some important features that merit discussion. Figure \ref{FIGURE:SpatialDistBlitzes} shows that many sections of Fortaleza street were treated between 2012 and 2013. In addition, places are repeatedly treated\footnote{The darker blue and purple circles and very close circles capture the overlapping of interventions around the same location over time.} within a year at different times, since the police command explicitly wants to give drivers the impression that there is always a risk of being stopped by a blitz.
			
			\noindent Hence, this quasi-random in space-time policing assignment aims to produce initial and residual deterrence. The sudden increase in the presence of police would generate initial deterrence by increasing the probability of apprehension of offenders, together with a potential ``free bonus'' associated with the lasting effects of blitzes, as criminals are not sure about future police operations in the area. In that sense, blitzes are similar to drunk-driving crackdowns in the United States.
			
			\noindent Table \ref{TABLE:BlitzesCharac} shows that these operations are easily seen and last from 30 minutes to 8 hours. The rationale behind limiting the time devoted to each target is that the police can avoid wasting scarce resources by minimizing the initial deterrence decay - as time passes, criminals get to know where cops are located and can avoid these street segments. Some parts of the city received more police interventions than others, but the blitzes are not driven by crime waves, as the main goal of the State Highway Police is to improve the safety of the roads. The Bureau of Operations met every Friday and decided the allocations of the blitzes in the subsequent week, considering their past allocation to create the optimal deterrence effect (initial plus residual). 
			
			\begin{table}[!h]
				\centering\small
				\caption{Blitzes characteristics in Fortaleza-CE, 2012-13}
				\label{TABLE:BlitzesCharac}
				\begin{tabular}{lccc}
					\\[-1.8ex]\hline 
					\hline \\[-1.8ex] 
					& Mean & Maximum & Minimum \\ 
					\hline \\[-1.8ex] 
					\rule{0pt}{3ex}
					Duration (hours)   & 5.57   & 8 & 0.5  \\ 
					\rule{0pt}{3ex}
					Number of policemen    & 5.84 & 75 & 2  \\ 
					\rule{0pt}{3ex}
					Police vehicles allocated     & 1.8 & 21 & 0  \\ 
					\\[-1.8ex]\hline 
					\hline \\[-1.8ex] 
				\end{tabular}%
				\label{tab:addlabel}%
				\\Source: Cear{\'a}'s Secretary of Public Security (SSPDS-CE).
			\end{table}%
			
			\noindent On average, five to six officers work 5.5 hours in a street segment, approaching around 96 and seizing about five vehicles per blitz. Weapons and drugs were found in 1\% of the interventions. The State Highway Police decides where to allocate blitzes within the State, how long to stay, and their composition, that is, the number of policemen and equipment used. Around 59\% of the blitzes have the support of police officers on motorcycles to catch drivers who turn to avoid the blitz. It is important to mention that the Military Police has other branches responsible for different policing assignments, and there is no significant deviation of resources to assemble blitzes. 
			
			\begin{figure}[!h]
				\centering 
				\subfloat[Blitz duration versus Vehicles Approached]{{\includegraphics[width=7cm]{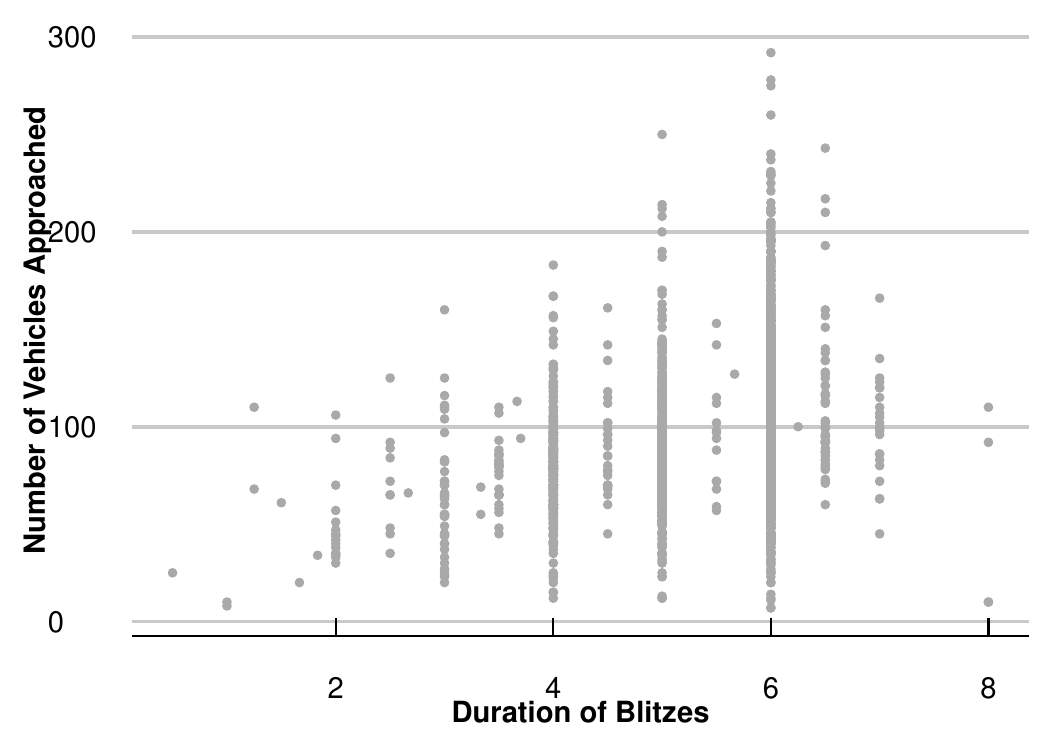}}}
				\quad
				\subfloat[Vehicles Approached versus Seized Vehicles]{{\includegraphics[width=7cm]{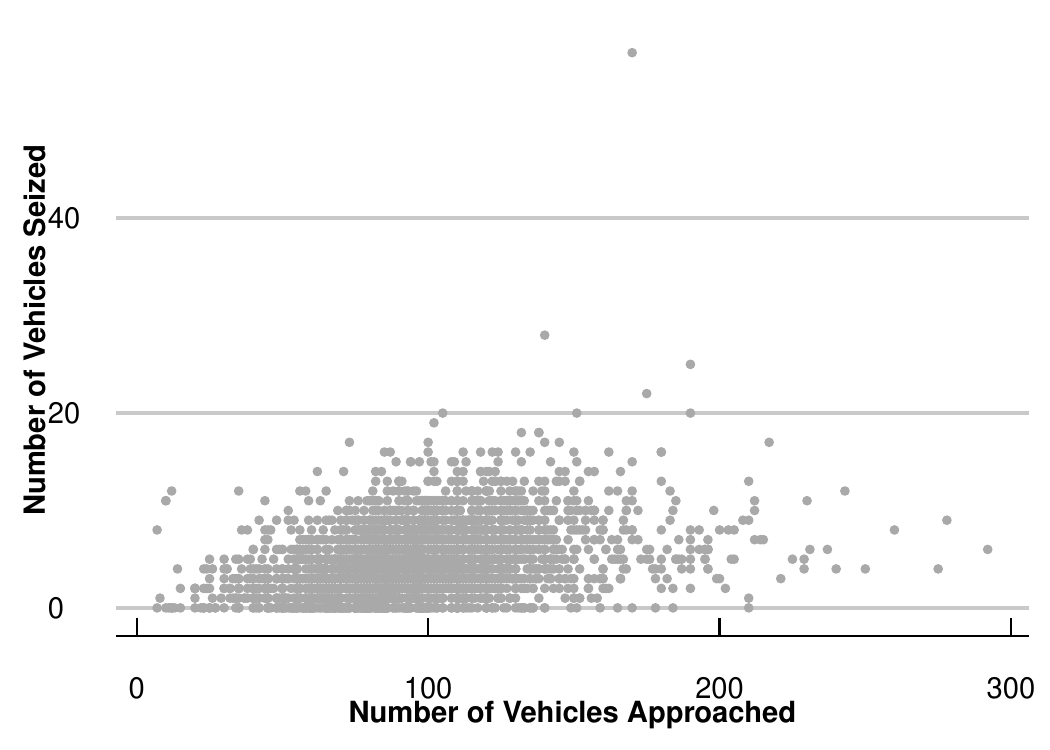}}}
				\caption{Outputs of Blitzes in Fortaleza-CE, 2012-2013}
				\label{FIGURE:OutputBlitzes}
			\end{figure}
			
\begin{figure}[!h]
				\centering 
				\subfloat[Fixed]{{\includegraphics[width=10cm, height=7cm]{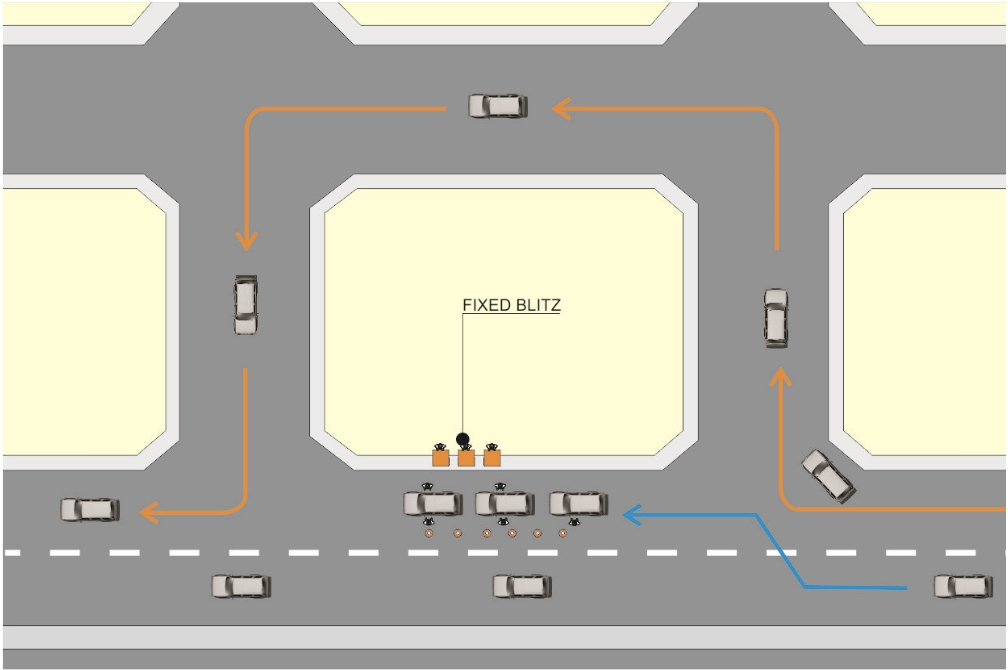}}}
				\quad
				\subfloat[Mobile]{{\includegraphics[width=10cm, height=7cm]{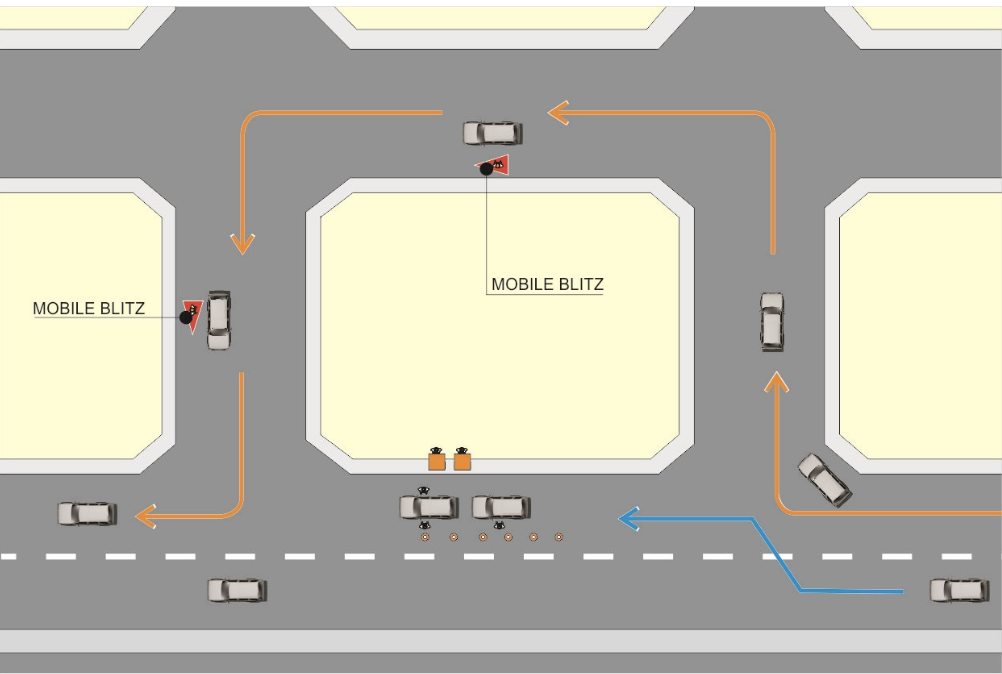}}}
				\caption{Blitzes Types}
				\label{FIGURE:TypeBlitzes}
			\end{figure}

			\noindent Figure \ref{FIGURE:OutputBlitzes} shows the correlations between the time spent in a blitz, the number of vehicles selected for inspections, and the number of seized vehicles.  It is worth mentioning that diminishing returns appear in both associations, which agrees with the idea of initial deterrence decay. After a couple of hours, drivers with potential traffic violations might be informed of the police intervention and avoid certain streets. When assembling a blitz, the command can opt for a wholly fixed assignment or set up the standard linear row of policemen with the additional support of police officers driving motorcycles around the blitz (Figure \ref{FIGURE:TypeBlitzes}). 
			
			\noindent The main advantage of mobile blitzes is that motorcycle officers can follow drivers who turn to avoid the police. It is fair to assume that these drivers are more likely to get traffic tickets or be arrested. However,  mobile blitzes stop fewer vehicles compared to fixed blitzes, as they divert some of their policemen to drive motorcycles. This means that there is a trade-off between quantity and ``quality'' of stopped and searched vehicles that can only be assessed empirically. This heterogeneity in terms of blitzes (multidimensional treatment) will be modeled and assessed later on.
			
			\subsection{Crime Data}
			
			The State Secretariat of Public Security (SSPDS-CE) shared records of all registered robberies (including violent crimes such as vehicle (carjacking) and cargo robberies, robberies against persons, and kidnapping, among others; when a robbery is followed by murder, the crime is accounted for as murder) and murders between 2012 and 2013.
			
			Most murder locations have street names and numbers and could be geolocated with high precision. However, some robberies have incomplete records, sometimes missing street numbers or both addresses and numbers.

			\begin{figure}[!h]
				\centering
				\subfloat[Murders]{{\includegraphics[width=7.1cm]{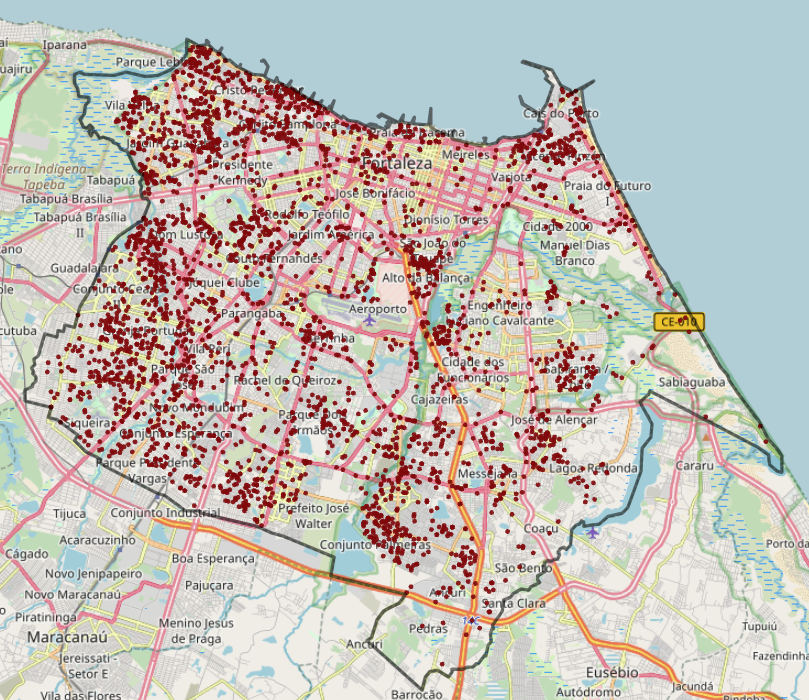}}}
				\quad
				\subfloat[Robberies]{{\includegraphics[width=7.03cm]{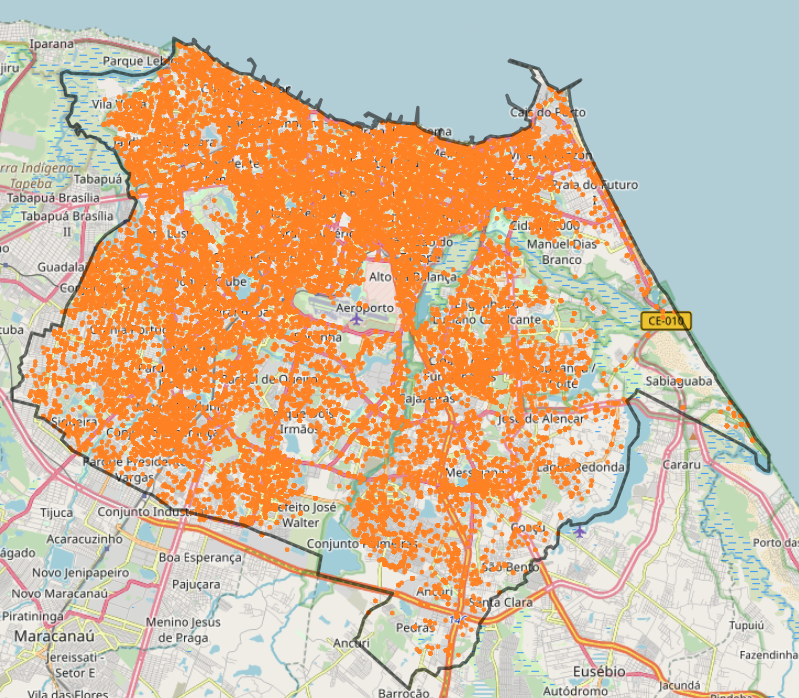}}}
				\caption{Spatial distribution of violent crime within Fortaleza, 2012-13.}
				\label{FIGURE:CrimeDistribution}
			\end{figure}

			\noindent In addition to addresses and type of occurrence, crime records also have the date and time of each event, see Figure \ref{FIGURE:CrimeDistribution}. Using Google Maps API, we could geolocate 69,243 violent crimes - 3,383 murders, and 65,860 robberies had all the information necessary to get relatively precise coordinates. Although robberies occur everywhere in the city, they tend to concentrate more in the center (north), while murders usually cluster in the city's periphery. 
			
			\subsection{Sample Construction}\label{SUBSECTION:Sample Construction}
			
\begin{figure}[!h]
				\centering
				\subfloat[Fortaleza's divided into 2,562 hexagons]{{\includegraphics[width=8.5cm, height=7.5cm]{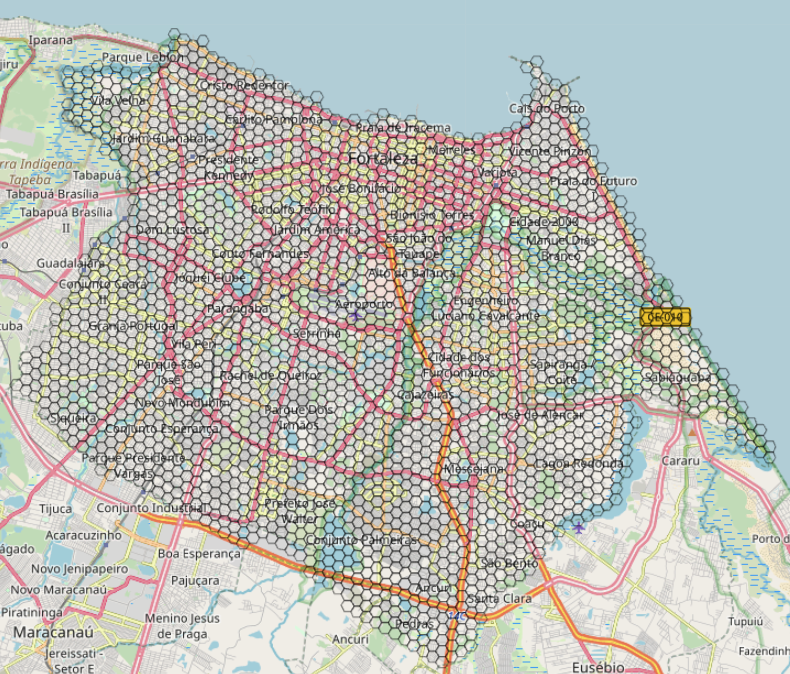}}}
				\quad
				\subfloat[Examples of cells in the Fortaleza's north region]{{\includegraphics[width=8.5cm, height=7.5cm]{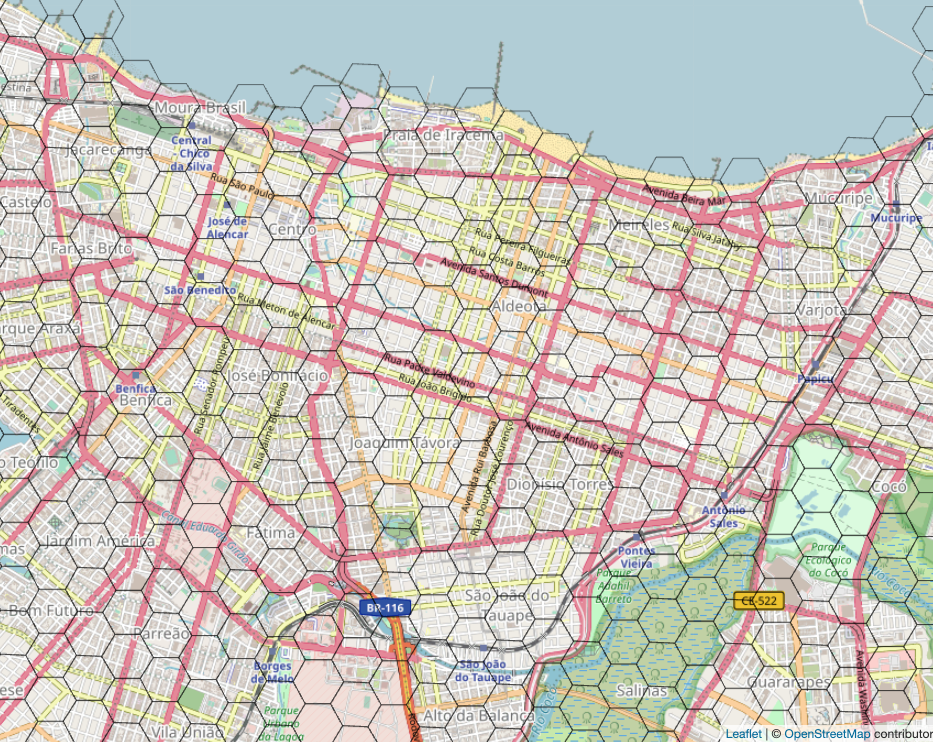}}}
				\caption{City’s territorial division.}
				\label{FIGURE:TerritorialDiv}
			\end{figure}

			We fill Fortaleza's area with a hexagonal spatial grid\footnote{The Institute for Applied Economic Research (Ipea) provides the shapefiles with spatial grids - see Ipea's Access to Opportunity Project.} with 2,526 cells, where each cell has an area of 0.126 $km^{2}$ (0.049 square miles), covers around 943 people, on average, and has a very similar size to regular census tracts as revealed in Figure \ref{FIGURE:TerritorialDiv}. In general, cells aggregate 3-6 fragments of street segments, and this high resolution allows the estimation of spatial spillovers and curbs minor measurement errors that naturally arise in georeferenced data.
			
			The days from January $1st$, 2012 to December $21^{th}$, 2013 are divided into four periods of six hours: morning, afternoon, night and dawn\footnote{The morning covers the hours from 6 am to 11:59 am, afternoon 12 pm until 5:59 pm, night 6 pm to 11:59 pm, and dawn from 12 am to 5:59 am.}, and a balanced panel with 7,388,808 rows\footnote{The 2,562 cells are observed over 2,884 day-time of day periods.} is constructed.
			
			\noindent Registered crime occurrences and hours of police work in blitzes\footnote{Some blitzes have a duration higher than six hours, and some start at a certain period of the day and stays until the next. In these cases, we assign the corresponding number of blitz hours within each period of the day. When two interventions are placed in the same cell at the same day-time of the day, we cap the number of policing hours to six.}\footnote{The sum of police work hours in the panel is higher compared to the raw data because we allocate a value of .5 to each 30-minute interval within the days between 2012 and 2013, even if a blitz spend less than 30 minutes in that time interval. For example, if a blitz starts or ends at 8:45 am, we still consider the interval between 8:30 and 9:00 am.  One can interpret this as a time that police officers need to set up or dismantle a blitz.} are located inside the cells and aggregated in the day-time-of-day dimension, with zeros filling the cells without any policing assignment and/or crime at a certain day-of-day. The outcome analyzed is violent crimes (the sum of robberies and murders), and the total number of hours spent by a blitz is the treatment.   
			
			Table \ref{tab:addlabel} shows the daily averages of crimes and hours of policing work in periods of the day. In fact, blitzes usually occur at night (between 6 pm and 11 pm), when most murders and robberies occur, and about 21.5\% of the cells were treated at some point between 2012 and 2013.
			
			\begin{table}[!h]
				\small
				\centering
				\caption{Policing and crime within time of day}
				\begin{tabular}{lcccccc}
					\\[-1.8ex]\hline 
					\hline \\[-1.8ex] 
					& 	\multicolumn{2}{c}{Murder} & 	\multicolumn{2}{c}{Robbery} & 	\multicolumn{2}{c}{Blitz (Hours)}  \\ 
					\hline \\[-1.8ex] 
					& Total & Mean & Total & Mean & Total & Mean \\ 
					\rule{0pt}{3ex}
					Dawn (12:00 am to 5:59 am) & 645   & 0.00043   & 5,983 & 0.00324 & 3,332 & 0.0018 \\ 
					\rule{0pt}{3ex}
					Morning (6:00 am to 11:59 am) & 478  & 0.00035 & 15,927 & 0.00862 & 1,962 & 0.0011 \\ 
					\rule{0pt}{3ex}
					Afternoon (12:00 pm to 5:59 pm)  & 788   & 0.00026  & 21,803 & 0.0118 & 2,489 & 0.0014 \\ 
					\rule{0pt}{3ex}
					Night (6:00 pm to 11:59 pm)  & 1,472   & 0.00080 & 22,147 & 0.0120 & 13,014 & 0.0071  \\ 
					\\[-1.8ex]\hline 
					\hline \\[-1.8ex] 
					\multicolumn{7}{l}{\footnotesize{Source: Ceará Secretary of Public Security (SSPDS-CE).}}\\
				\end{tabular}%
				\label{tab:addlabel}
			\end{table}

			\noindent The high resolution of the data provides a unique opportunity to identify the causal effects of local police interventions, but also presents some estimation challenges. For example, 99.1\% of the outcomes are zero, which complicates the use of linear models. The following section discusses how we leverage the intervention's design and the information available for identification, and details the estimation method adopted.   
			
			\section{Research Design}\label{SECTION:ResearchDesign}
			
			\subsection{Identification}
			
			As mentioned above, it is challenging to isolate the causal effect of police on crime because crime levels frequently determine variations in police presence, which misleads the relationship between crime and policing to be positive. However, when assembling blitzes, the Police command aims to surprise drivers in space-time and give them the impression that they could be subject to stop and search at any time. 
			
			By the same token, personal conversations with senior officers responsible for blitzes allocation add confidence in a 'quasi-natural experiment' hypothesis, which we implicitly incorporate into our econometric identification strategy later on. Those officers reported that the weekly spatial allocation of the blitzes followed an almost personal random pattern where they chose locations (as well as the intensity and duration of the blitzes) considering the strategic response of potential criminals\footnote{Although randomization by humans is still an open debate, see \citeonline{ Towse1998-hs} and \citeonline{Wong2021-cm},  the human ability to generate random sequences (RSG) improves with feedback, and many scientific results demonstrate that human RSG can reach levels statistically indistinguishable from pseudorandom computer.}. Hence, by design, this policing assignment does not react to past crime waves, alleviating concerns about simultaneity usually present in the literature. In addition, the high spatial resolution and high time frequency create additional opportunities to isolate the direct and indirect causal effects of policing on crime.  
			
			To estimate policing spillovers, we build 2,562 by 2,562 spatial weight matrices based on the inverse distance of cells' centroids to each other. The elements of these matrices are defined as $w_{ij}=1/d_{ij}$, where $d_{ij}$ is the distance between the hexagon $i$ and the hexagon $j$. The rationale behind this approach is that the effects of policing decrease with increasing distance. From the offender's perspective, it is straightforward to presume limitations to mobility within the city and a preference for familiar streets. For that reason, we also establish cutoffs up to 1.5 km since assuming place-based interventions producing spillovers in the entire city's area is, at least, heroic. Therefore, blocks up to 500-1500 meters from a hexagon $i$ have weight $1/d_{ij}$, while blocks further away have 0 weight and do not influence crime occurrences in cell $i$. 
			
\begin{figure}[!h]
				\centering
				\subfloat[]{{\includegraphics[width=7cm]{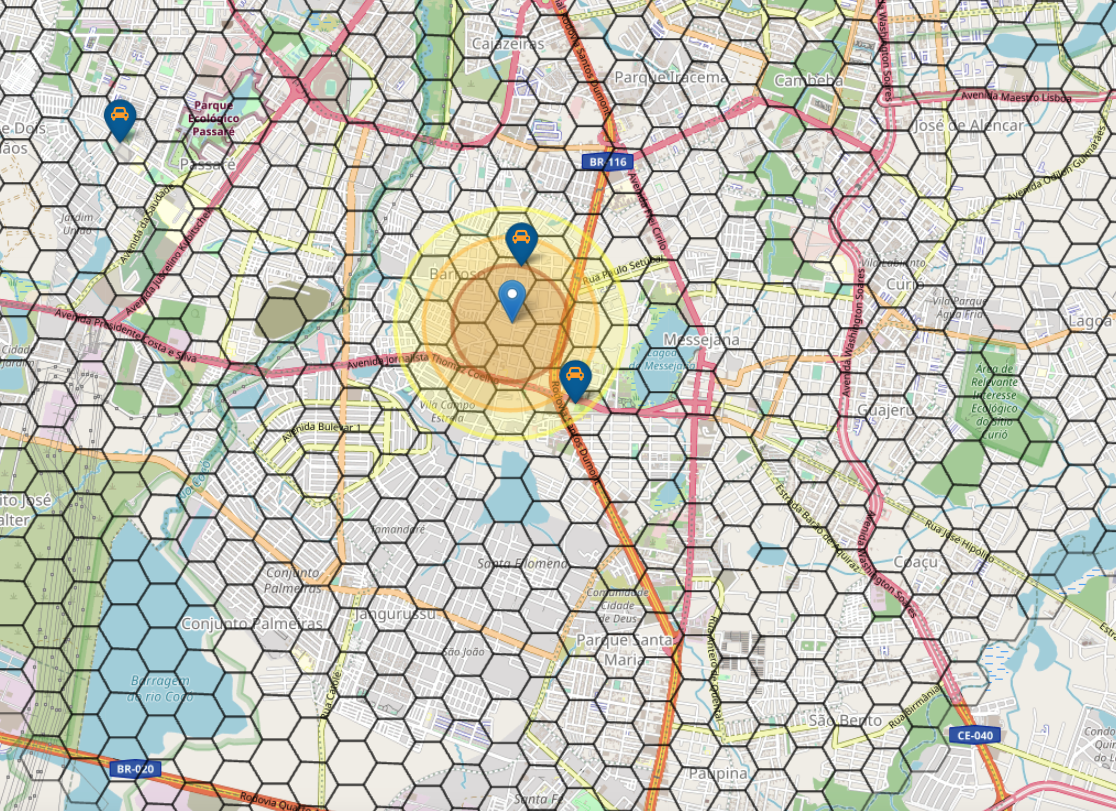}}}
				\quad
				\subfloat[]{{\includegraphics[width=7cm]{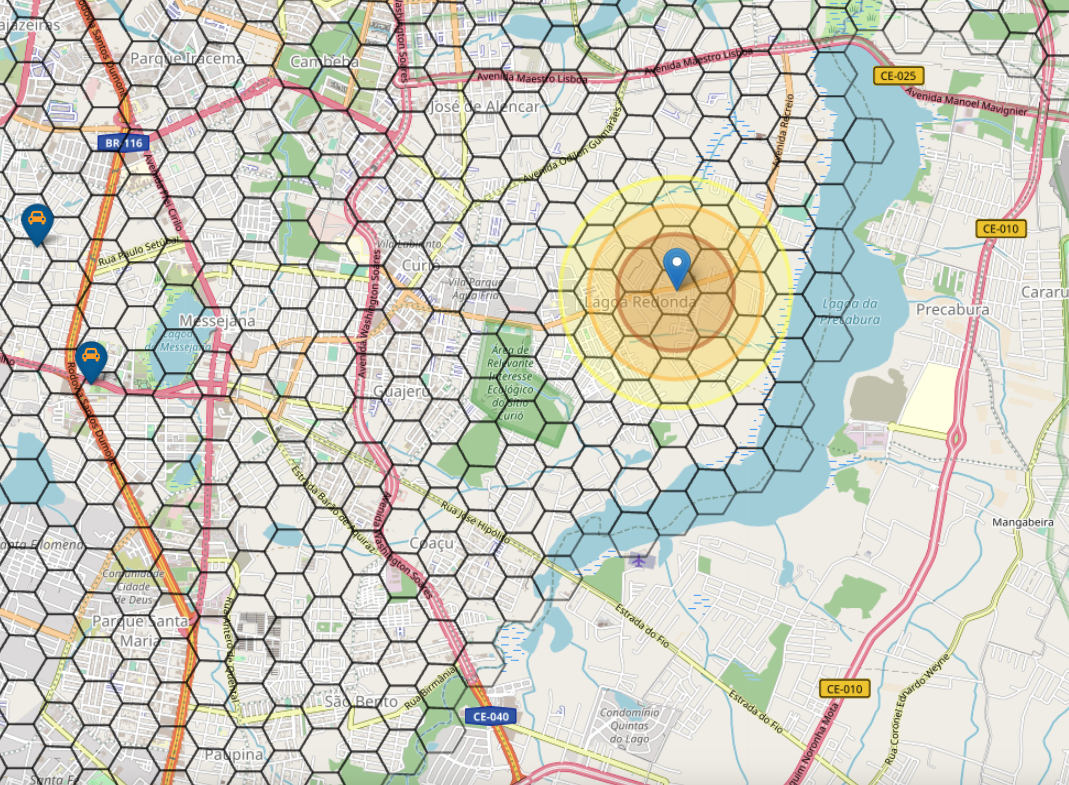}}}
				\caption{Blitz allocation during January 9$^{th}$, 2012}
				\label{FIGURE:BlitzAllocationEX}
			\end{figure}

			Figure \ref{FIGURE:BlitzAllocationEX} illustrates the empirical strategy for dealing with spatial spillovers. The yellow circle has a radius of 1 km. 
			
			\noindent In contrast, orange and red circles have 500 and 250 m radii, respectively, and these are the influence areas of a blitz in the neighborhood of a given cell, represented by the light blue balloon. The hexagons with centroids placed inside the circle will have a weight of $1/d_{ij}$, and the others 0. This weight matrix (\textbf{W}) pre-multiply the vector of the daily number of hours spent by the police in a specific cell at a given time of the day to determine the weighted average number of policing hours in the surrounding area of a typical hexagon.
			
			In addition to the contemporaneous spatial effects of local police interventions in neighboring regions, another possible causal channel of blitzes and crime is related to residual deterrence (diffusion) and temporal displacement, see \citeonline{sherman1990police}. As stated earlier, one of the goals of this policing assignment is to combine the short-term effects of the presence of police on crime in a given area with potentially lasting effects over the following days by surprising drives in space and time with small but frequent interventions. 
			
			On the other hand, crime can be moving around the clock. In this case, the offender would wait until the blitz is gone to go back to business as usual. Local police interventions would only delay criminal activity, as police cannot always be everywhere. To capture residual deterrence and temporal displacement, the reduced-form model incorporates part of the history of blitzes with temporal lags of hours spent by police in the past days (distributed-lag model).     
			
			Although simultaneity issues are unlikely given the design of the blitzes\footnote{Both linear and Poisson fixed effects regressions of the past number of crime occurrences on hours of policing estimates null results for all 16 lags (past four days) of crime. The results are available in the online appendix}, there might be some characteristics of the street segment that are correlated with crime and influence the allocation of blitzes. For example, commanders might assign interventions to areas with high vehicle traffic, which could be areas targeted by criminals. To mitigate omitted variable bias, we add cells-by-day of the week-by-time of day and daily fixed effects. 
			
			Effectively, we compare crime and policing hours in hexagons with themselves on the same day of the week at the same period of the day during 2012 and 2013 while controlling for common daily effects. Hence, the main identification assumption is conditional independence. Given this rich set of fixed effects, the hours spent by a blitz and its spatial and temporal lags are exogenous in the police crime regression.

			\subsection{Estimation}
			
			We start by formally testing first the diminishing returns and the effects of blitz type on blitzes outcomes, before the econometrics of impact evaluation. We regress the number of vehicles stopped on blitz duration, and the number of traffic tickets and vehicles seized on the number of stops.
			
			Table \ref{TABLE:EstimationDiminishingReturns} shows the results of fixed effects\footnote{Cell is the unit of observation adopted in the paper, see Subsection \ref{SUBSECTION:Sample Construction}.} linear regressions. On average, one additional hour of a blitz is associated with 29 more vehicles approaching. In addition, there are one additional seizure and four additional tickets for an increase of 25 vehicle inspections. To maximize vehicle stops, the blitz duration should be around 6 hours, close to the sample's average of 5.57. It is important to stress the fact that the commander of the blitz (always a more experienced high-rank police officer) has total discretion as to when to shut down the operation. As we shall see, police discretion is a highly debated and scientifically significant topic. Another important result is that mobile blitzes are more successful in distributing traffic tickets and seizing vehicles. On average, a mobile blitz increases traffic tickets and vehicle seizures by 10.7 and 16.5\%, respectively. 
			
			\begingroup
			\begin{table}[!h]
				\footnotesize
				\centering
				\caption{Fixed Effects Linear Regression Results - Blitzes Outcomes}
				\label{TABLE:EstimationDiminishingReturns}
				\begin{tabular}{lccc}
					\tabularnewline \midrule \midrule
					Dependent Variables:         & Vehicles Stopped   & Traffic Tickets & Seized Vehicles\\   
					& (1)                   & (2)\\  
					\midrule
					Blitz Duration                     & 31.84$^{***}$         &                         &   \\   
					& (8.081)               &                         &   \\   
					Blitz Duration$^2$              & -2.636$^{***}$        &                         &   \\   
					& (0.8516)              &                         &   \\   
					Number of Officers                   & 1.218$^{***}$         & 0.3056$^{***}$          & 0.2457$^{**}$\\   
					& (0.3715)              & (0.0986)                & (0.1172)\\   
					Blitz Type Mobile                       & 2.153                 & 1.767$^{***}$           & 0.7897$^{***}$\\   
					& (2.271)               & (0.5541)                & (0.2473)\\   
					Vehicles Stopped        &                       & 0.1553$^{***}$          & 0.0447$^{***}$\\   
					&                       & (0.0159)                & (0.0066)\\   
					Vehicles Stopped$^2$  &                       & -0.0003$^{***}$         & $-8.63\times 10^{-5 ***}$\\    
					&                       & ($7.14\times 10^{-5}$)  & ($2.96\times 10^{-5}$)\\   
					\midrule
					\emph{Fixed-effects}\\
					Cell                      & Yes                   & Yes & Yes\\  
					Day                  & Yes                   & Yes & Yes\\  
					Period of the Day                & Yes                   & Yes & Yes\\  
					Day of the Week            & Yes                   & Yes & Yes\\  
					\midrule
					Outcome Mean               & 95.9    & 16.5    & 4.79\\
					Observations                 & 3,409                 & 3,409 & 3,409\\  
					R$^2$                       & 0.48693               & 0.65483                 & 0.60094\\  
					\midrule \midrule
					\multicolumn{4}{l}{\emph{Standard-errors clustered at cell in parentheses}}\\
					\multicolumn{4}{l}{\emph{Signif. Codes: ***: 0.01, **: 0.05, *: 0.1}}\\
				\end{tabular}
			\end{table}
			\par\endgroup
			
			\noindent Now, to measure direct and indirect impacts of blitzes on violent crime occurrences, the following reduced-form model is estimated.
			
			\begin{equation}
				\label{EQ:Impact01}
				ln(\lambda_{idt}) = \delta Blitz_{idt}+\theta Blitz^{2}_{idt} +\rho WBlitz_{idt}+ \Gamma lag(Blitz_{idt}, dt-j) +c_{ipw}+ \alpha_{d} +  \varepsilon_{idt} \\
			\end{equation}
			
			We assume that $\text{Crime}_{idt} \sim \text{Poisson} (\lambda_{idt})$. The use of the Poisson distribution is justified because the dependent variable $\text{Crime}_{idt}$ is the number of crime occurrences (count variable) measured in the cell $i$ and the daytime of the day $dt$. $Blitz_{idt}$ refers to the number of hours spent by police officers in cell $i$ at day-time of day $dt$, $WBlitz_{idt}$ is the weighted average of blitzes hours in the surrounding area of hexagon $i$ at day-time of day $dt$, and $lag(Blitz_{idt}, dt-j)$ represents the hours of policing work at cell $i$ at day-time of day $dt-j$\footnote{For instance, if $dt$ refers to the afternoon of January 9$^{th}$, 2012, $dt-1$ is the morning of January 9$^{th}$, 2012, $dt-4$ is the afternoon of January 8$^{th}$, 2012, and so on.}. In addition, $c_{ipw}$ means fixed effects by period of the day by day of the week, $\alpha_{d}$ is the intercept that varies every day, and $\varepsilon_{idt}$ represents the error term. Finally, we introduce the quadratic term in Equation \ref{EQ:Impact01}, assuming diminishing returns of policing hours in the same area.  
			
			As we do not have a clear economic theory or precise mechanism for the spillover effects, and as we are dealing with nonexperimental data, we follow \citeonline{halleck2015slx} advice and use as a benchmark their spatial lag of \textbf{X} model (SLX). In addition, the advantage that the spillover effects using the SLX model are more straightforwardly applied, both in terms of estimation and interpretation, they are also more flexible than traditional choices in the spatial econometrics literature. 
			
			Time-varying intercepts capture common daily trends across cells, such as the wet season. Grids have approximately 0.12 $km^{2}$, and these fixed effects interacted with day-of-week and time-of-day control for unobserved heterogeneity in small areas on different days of the week and periods of the day. For example, vehicle traffic within cells probably has regular patterns at specific periods of the day during certain days of the week. Although we do not observe these patterns in the data, the model takes these into account with a rich set of fixed effects and includes spatial\footnote{The inclusion of the spatial lag is similar to what \citeonline{halleck2015slx} call SLX model.} and temporal lags to capture displacement effects.
			
			The parameters of interest are $\delta$ and $\theta$, which measure the average direct effects of local interventions, $\rho$, which capture the effects of blitzes in surrounding regions (displacement or spillover effects), and $\Gamma$, a vector of parameters associated with the history of blitz deployment that captures residual deterrence and temporal displacement. 
			
			\section{Results}\label{SECTION:Results}
			
			\subsection{Direct and Indirect Effects of Local Police Interventions on Crime}
			
			Table \ref{TABLE:MainEstimatives} shows the direct and indirect (spatial and temporal) effects of local police interventions considering temporal lags up to 16 day-periods per day, i.e. up to four days after an intervention\footnote{Results remain very similar across specifications that consider temporal lags from 4 to 28 - previous 24 hours to a week from the policing assignment.}. Blitzes have a meaningful and statistically significant direct effect on violent crime occurrences, and an additional hour spent by this policing assignment causes an average decrease of 21\% on daily crime counts in cells at a given 6-hour period of day\footnote{That is, $(e^{-0.281+0.046}-1) \times 100$.}. 
			
			To estimate the spatial effect of blitzes, we consider a range of different spatial weight matrices (\textbf{W}). The first column contains the estimated parameters of Equation \ref{EQ:Impact01} when considering all hexagons that share a border with a given cell, placing the same weights on each of the six neighbors\footnote{Cells located at the border of the city might have only 2 or 3 neighbors.}. Columns (2) to (5) use rings of 500-1500 meters in radius to determine catchment areas, placing higher weights on closer neighbors, as discussed in the previous section. On average, the 500-meter ring encircles 5.8 cells, while the 750-, 1000-, and 1500-meter rings encompass 11.46, 17.1, and 49.5 cells. 
			
			In the preferred specification using an inverse distance spatial weight matrix with a 1 km cutoff,  $\hat{\rho}$ points to an average decrease of 0.3\%\footnote{That is, $(e^{-0.0529\times \frac{1}{17.1}}-1) \times 100$. This is an approximation since the weights are higher (lower) for closer (farther) interventions.} in violent crime counts in a given cell followed by an additional hour of policing in its neighboring region, which is small and not statistically different from zero. 
			
			All in all, there is no evidence of spatial effects across a wide range of weight matrices: neither a diffusion of benefits of blitzes to immediate neighbors nor crime displacement to nearby cells\footnote{Overall, coefficients related to temporal lags of spatially lagged blitz hours, i.e., $lag(WBlitz_{idt}, dt-j)$, are not statistically different from zero as a group and do not change estimates presented in Table \ref{TABLE:MainEstimatives}. Hence, there are no initial or lasting spatial effects.}. In addition, most of the parameters associated with the previous blitz deployment ($\Gamma$) are indistinguishable from zero, and Table \ref{TABLE:MainEstimatives} shows the ones statistically significant at the usual levels. In the next six hours after a local intervention, violent crime incidents appear to increase by 6.6\%, on average. On the other hand, average decreases in crime around 5.9 and 4.5\% are observed 42 and 72 hours after a blitz. These results suggest some strategic interactions between criminals and the police force. Although the Police want to enforce a reputation of fighting crime relentlessly, criminals may find it unlikely since officers do not return to the same place in the next couple of hours from an intervention. However, offenders might expect them to return to the same area after 2 or 3 days, given intermittent blitz assignments. 
			
			Regarding overall lasting effects, we could not reject the null of no joint significance of lags up to 16. Hence, we do not find residual deterrence or temporal displacement associated with blitzes. We observe neither spatial nor temporal displacement of crime, and direct effects are statistically indistinguishable from total effects. The intervention, by design, places officers in different areas at distinct times, and even though it does not generate residual deterrence, the assignment's uncertainty might help minimize temporal displacement. 
			
			Another interesting result is the diminishing returns of public safety to hours spent by police in a single area, as $\hat{\theta}$ shows a relevant effect, which agrees with the decreasing marginal productivity of blitzes duration on the number of vehicles inspected. Blitzes that spend too many hours in the same place lose the element of surprise, and the flow of information about the location of blitzes most likely makes drivers use different streets to commute. In that sense, the optimal amount of time a blitz should spend in an area in a given 6-hour period to minimize violent crime is 3.04. However, this policing assignment might have other goals, such as minimizing drunk driving and maximizing the distribution of traffic tickets. 
			
			\begin{landscape}
				\begingroup
				\begin{table}
					\small
					\centering
					\caption{Fixed Effects Poisson Regression Results}
					\label{TABLE:MainEstimatives}
					\begin{tabular}{lccccc}
						\tabularnewline \midrule \midrule
						& \multicolumn{5}{c}{\textbf{Dependent Variable: Violent Crimes Occurrences}}\\
						& Binary Contiguous            &  500 meters            & 750 meters           & 1 km           & 1.5 km \\
						& (1) & (2) & (3) & (4) & (5) \\
						\midrule
						$Blitz$               & -0.2814 & -0.2815 & -0.2809 & -0.2808 & -0.2809\\ 
						& (0.1360)$^{**}$       & (0.1360)$^{**}$       & (0.1360)$^{**}$       & (0.1360)$^{**}$       & (0.1360)$^{**}$\\
						&     & \{0.1271\}$^{**}$       & \{0.1305\}$^{**}$       & \{0.1221\}$^{**}$       & \{0.1376\}$^{**}$       \\[.1cm]
						$Blitz^{2}$              & 0.0463   & 0.0463  & 0.0462   & 0.0461   & 0.0461\\
						& (0.0250)$^{*}$       & (0.0250)$^{*}$        & (0.0250)$^{*}$       & (0.0250)$^{*}$       & (0.0250)$^{*}$\\ 
						&   & \{0.0236\}$^{*}$        & \{0.0240\}$^{*}$       & \{0.0224\}$^{**}$       & \{0.0254\}$^{*}$       \\[.1cm]
						$WBlitz$                    & 0.0072         &  0.0085            & -0.0413          &  -0.0529              &-0.0816   \\   
						& (0.0611)       & (0.0635)           &  (0.0885)        &  (0.1046)              & (0.1760)  \\ 
						&     & \{0.0701\}       & \{0.0924\}       & \{0.1213\}       & \{0.2398\}       \\[.1cm]
						lag($Blitz$,1)      & 0.0638  & 0.0638  & 0.0639  & 0.0638  & 0.0638\\ 
						& (0.0295)$^{**}$       & (0.0295)$^{**}$       & (0.0295)$^{**}$       & (0.0295)$^{**}$       & (0.0295)$^{**}$\\ 
						&        & \{0.0319\}$^{**}$       & \{0.0344\}$^{*}$       & \{0.0341\}$^{*}$       & \{0.0301\}$^{**}$        \\[.1cm]
						lag($Blitz$,7)          & -0.0609 & -0.0609 & -0.0609 & -0.0609 & -0.0609\\  
						& (0.0292)$^{**}$       & (0.0292)$^{**}$       & (0.0292)$^{**}$       & (0.0292)$^{**}$       & (0.0292)$^{**}$\\
						&    & \{0.0298\}$^{**}$         & \{0.0286\}$^{**}$       & \{0.0313\}$^{*}$       & \{0.0345\}$^{*}$ \\[.1cm]
						lag($Blitz$,12)         & -0.0460  & -0.0460  & -0.0459  & -0.0459  & -0.0459\\  
						& (0.0240)$^{*}$       & (0.0240)$^{*}$       & (0.0240)$^{*}$       & (0.0240)$^{*}$       & (0.0240)$^{*}$\\ 
						&   & \{0.0249\}$^{*}$       & \{0.0238\}$^{*}$       & \{0.0254\}$^{*}$       & \{0.0236\}$^{*}$      \\
						\midrule
						\emph{Fixed-effects}\\
						Cell-by-time of day-by-day of week    & Yes            & Yes            & Yes            & Yes            & Yes\\  
						Daily                & Yes            & Yes            & Yes            & Yes            & Yes\\  
						\midrule
						Observations                & 2,608,265      & 2,608,265      & 2,608,265      & 2,608,265      & 2,608,265\\  
						BIC                         & 970,485.1      & 970,485.1      & 970,484.9      & 970,484.8      & 970,484.9\\  
						\midrule \midrule
						\multicolumn{6}{l}{\emph{Clustered (cell) standard-errors in parentheses, and Conley's standard-errors in brackets (500-1500 meters,}}   \\
						\multicolumn{6}{l}{\emph{according to the distance cutoffs of weight matrices). Note: temporal lags up to 16 day-time of day periods (4 days}}\\
						\multicolumn{6}{l}{\emph{from an intervention) are included, but only statistically significant ones are shown. ***: 0.01, **: 0.05, *: 0.1}}\\

					\end{tabular}
				\end{table}
				\par
				\endgroup
			\end{landscape}

			\subsection{Mechanisms and Heterogeneity}
			There are two main mechanisms through which these local police interventions could affect crime occurrences: deterrence and incapacitation \citeonline{Kessler1999-ef}. The fact that police officers stay in the same segment of the street for more than five hours on average might induce a flow of information about the allocation of police in a certain area. It is important to emphasize that blitzes are very visible operations, which might immediately affect the offender's perceived risk of apprehension, increasing the cost of committing a crime. Another possibility is that this specific policing assignment catches drivers who target that area for criminal activities, which would induce incapacitation effects. 
			
			To partially unravel these mechanisms, we explore information about the number of motor vehicles and motorcycles seized during each intervention\footnote{The data also have the number of weapons collected, but only 28 out of 3,423 interventions recorded at least one type of weapon as an output of the operation.}. Vehicle apprehension usually occurs when a driver does not have a license or vehicle registration or appears to be under the influence of alcohol or illegal substances, and cannot designate another person to drive the vehicle. When the Police find weapons and drugs, vehicle occupants are also taken to the police station. Although not all vehicle apprehensions are related to robbers and murderers, we believe that vehicle seizures are a reasonable measure of incapacitation.

			\begingroup
			\begin{table}[!h]
				\footnotesize
				\centering
				\caption{Fixed Effects Poisson Regression Results - Heterogeneity}
				\label{TABLE:EstimationDetIncap}
				\begin{tabular}{lccc}
					\tabularnewline \midrule \midrule
					\multicolumn{4}{c}{\textbf{Dependent Variable: Violent Crime Occurrences}}\\
					& 1 km          &  1 km           &  1 km           \\
					& (1) & (2) & (3) \\
					\midrule
					$Blitz$                 & -0.491 & -0.485  & -0.309 \\   
					& (0.161)$^{**}$        & (0.204)$^{**}$       & (0.136)$^{**}$       \\      
					& \{0.164\}$^{**}$       & \{0.195\}$^{**}$       & \{0.137\}$^{**}$   \\ 
					\text{Vehicles Seizure}           & 0.195        &       & \\   
					& (0.1151)       &        & \\   
					&  \{0.127\}    &       &    \\
					$Blitz$\text{:Vehicles Seizure}        & -0.003         &         &     \\   
					& (0.032)       &        &    \\   
					& \{0.165\}    &       &       \\
					\text{Policemen}                    &        & 0.166        & \\   
					&        & (0.120)       & \\   
					&     & \{0.111\}      &     \\
					$Blitz$\text{:Policemen}        &        & 0.002         &      \\   
					&     & (0.045)       &      \\   
					& & \{0.050\}       &     \\
					\text{Motorcycles}           &       &        & 0.722\\   
					&        &        & (0.529)\\   
					&     &     & \{0.506\}    \\
					$Blitz$\text{:Motorcycles}        &         &          &-0.114      \\   
					&        &        & (0.089)     \\   
					&        &        & \{0.088\} \\
					\midrule
					\emph{Fixed-effects}\\
					Cell x TOD x DOW  & Yes            & Yes            & Yes  \\ 
					Day                & Yes            & Yes            & Yes \\  
					\midrule
					Observations                & 2,608,265      & 2,608,265      & 2,608,265   \\  
					BIC                                           & 970,514.4      & 970,988.3      & 970,971.7 \\  
					\midrule \midrule
					\multicolumn{4}{l}{\emph{Clustered (cell) standard errors in parentheses, and Conley's standard}} \\
					\multicolumn{4}{l}{\emph{errors in brackets (1000 meters, according to the distance cutoff of the}}\\
					\multicolumn{4}{l}{\emph{weight matrix). Note: temporal lags up to 16 day-time of day periods}}\\
					\multicolumn{4}{l}{\emph{(4 days from an intervention) are included, but only statistically signi-}}\\   
					\multicolumn{4}{l}{\emph{ficant ones are shown. ***: 0.01, **: 0.05, *: 0.1 }}\\   
				\end{tabular}
			\end{table}
			\par\endgroup

			If incapacitation is the main channel through which blitzes reduce violent crime occurrences, we would see larger effects for blitzes that seize more vehicles. Table \ref{TABLE:EstimationDetIncap} shows the heterogeneous treatment effects of policing hours by the number of vehicles seized using the exact specification as in Table \ref{TABLE:MainEstimatives}. That means, we include $Blitz^{2}$ and both spatial and temporal lags (up to 4 days). As can be seen, the interaction of blitz hours with the number of vehicles collected is small and not statistically significant. Hence, the main mechanism is deterrence: by increasing the cost of committing a crime with visible operations, the police prevent crime occurrence.  
			
			Other potential sources of heterogeneity are related to the characteristics of the policing assignment. As mentioned above, blitzes can have different numbers of officers and can be fixed or mobile (see Section 3.2). In columns 2 and 3, Table 7, we interact the number of policemen and motorcycles\footnote{We also built a dummy that takes on one if the blitz has motorcycles (i.e., mobile blitz) and zero if the blitz is fixed (no motorcycle). The results remain very similar to the ones presented in Table 7.} assigned with the number of hours spent by the blitz in a given cell. We see that there is no significant direct impact of additional officers assigned to a blitz on crime, and mobile blitzes increase the direct deterrent effects by 11\%, but this impact is not statistically significant. We also observe residual deterrence for about 36 hours after a mobile blitz, which fades over in the following days. Overall, we cannot reject the null of joint significance of lags up to 16 day-time periods (after 4 days of an intervention).       
			
			\subsection{Robustness to Grid Size}
			
			To assess the robustness of the results to the cells' size, we construct a panel with a higher resolution, keeping the same time division (721 days, each with 4 periods) but dividing the city into 17,544 hexagons of 0.0179 square km\footnote{The H3 geospatial indexing system partitions the world into hexagonal cells, and one can choose a resolution that gives a certain number of hexagons inside the selected boundaries. The resolution equal to 10 gives the 17,544 hexagons mentioned, while the resolution equal to 9 gives 2,504, approximately the same number of cells in the shapefile provided by IPEA.} - a balanced panel with 50,596,896 observations. 
			
			\begin{table}[H]
				\footnotesize
				\centering
				\caption{Fixed Effects Poisson Regression Results - Higher Spatial Resolution}
				\label{TABLE:RobustnessCheck}
				\begin{tabular}{lcccc}
					\tabularnewline \midrule \midrule
					& \multicolumn{4}{c}{\textbf{Dependent Variable: Violent Crime Occurrences}}\\
					&  500 meters            & 750 meters           & 1 km           & 1.5 km \\
					& (1) & (2) & (3) & (4) \\
					\midrule
					$Blitz$                & -0.5242                & -0.5243              & -0.5241               & -0.5242\\ 
					& (0.2587)$^{**}$        & (0.2587)$^{**}$      & (0.2586)$^{**}$       & (0.2587)$^{**}$\\                         
					&  \{0.2178\}$^{**}$    & \{0.2217\}$^{**}$    & \{0.2194\}$^{**}$     & \{0.2159\}$^{**}$       \\[.1cm]
					$Blitz^{2}$            & 0.0853                 & 0.0854               & 0.0853                & 0.0853\\
					& (0.0469)$^{*}$        & (0.0469)$^{*}$       & (0.0469)$^{*}$        & (0.0469)$^{*}$\\ 
					&  \{0.0388\}$^{**}$     & \{0.0394\}$^{**}$     & \{0.0389\}$^{**}$     & \{0.0385\}$^{**}$       \\[.1cm]
					$WBlitz$               &  -0.1888               & -0.1724              & -0.4651               &-0.6030   \\   
					& (0.3662)               &  (0.5557)            &  (0.7327)             & (1.050)  \\ 
					& \{0.4044\}             & \{0.6206\}           & \{0.7964\}            & \{1.2748\}       \\[.1cm]
					lag($Blitz$,7)         & -0.0844                & -0.0844              & -0.0844               & -0.0844\\  
					& (0.0442)$^{*}$         & (0.0442)$^{*}$       & (0.0442)$^{*}$        & (0.0442)$^{*}$\\
					& \{0.0444\}$^{*}$      & \{0.0422\}$^{**}$    & \{0.0406\}$^{**}$      & \{0.0413\}$^{**}$ \\[.1cm]
					\midrule
					\emph{Fixed-effects}\\
					Cell x TOD x DOW               & Yes            & Yes            & Yes            & Yes\\
					Day                         & Yes            & Yes            & Yes            & Yes\\  
					\midrule
					Observations                  & 3,958,707      & 3,958,707     & 3,958,707      & 3,958,707\\  
					BIC                            & 1,244,904.6      & 1,244,904.9      & 1,244,904.5      & 1,244,904.6\\  
					\midrule \midrule
					\multicolumn{5}{l}{\emph{Clustered (cell) standard-errors in parentheses, and Conley's standard-errors }}   \\
					\multicolumn{5}{l}{\emph{in brackets (500 to 1500 meters, spherical distances, according to the distance }}\\
					\multicolumn{5}{l}{\emph{cutoffs of weight matrices). Note: temporal lags up to 16 day-time of day periods }}\\
					\multicolumn{5}{l}{\emph{(4 days from an intervention) are included, but only statistically significant }}\\
					\multicolumn{5}{l}{\emph{ones are shown. ***: 0.01, **: 0.05, *: 0.1. TOD is the time-of-day and DOW}}\\
					\multicolumn{5}{l}{\emph{represents the day-of-week.}}\\

				\end{tabular}
			\end{table}
			
			Table \ref{TABLE:RobustnessCheck} displays the results using the former model at this new level of aggregation. Direct effects are greater: an additional hour of police blitzes decreases violent crime occurrences by 35.5\%, on average. We observe the same pattern of diminishing returns of policing hours to public safety, no spatial displacement to nearby areas\footnote{Although the estimated impacts of $WBlitz$ are higher, spatial effects are very similar to the ones presented in Table \ref{TABLE:MainEstimatives} due to the smaller sizes of the cells - the higher resolution means that each distance cutoff covers more neighbors.}, and very little residual deterrence after two days of an operation.
			
			\subsection{Cost-benefit Analysis}
			
			In general, blitzes spend 3.28 hours in a cell in a given 6-hour period, which decreases violent crime occurrences by -34.62\%, on average. When blitzes are present in the 548 ever-treated cells, the daily average number of violent crimes at a period of the day is 0.0197. In the counterfactual scenario with no policing assignment, average violent crime occurrences would reach 0.0301. Therefore, one blitz prevents 0.0104 violent crimes from occurring in the first place in a given cell at a certain period of the day. 
			
			Between 2012 and 2013, these 548 cells had the presence of a blitz in 6,298 6-hour periods of the day\footnote{As mentioned before, many of the 3,423 interventions were present in two different periods of the day. For example, a blitz that started at 4 p.m. and ended at 9 p.m. is recorded as having spent 2 hours in the afternoon and 3 hours in the night}, and the total number of prevented violent crimes was approximately 66. Additional policing hours have similar effects on robbery and murder\footnote{Although estimates using murders are imprecise and not statistically significant at usual levels.}, and murders correspond to 5\% of all violent crimes. To estimate the social benefits regarding public safety, we follow \citeonline{pereira2020valor} and \citeonline{de2019assessing}, who calculate the value of a statistical life around R\$ 1.119 million for a Brazilian blue-collar man and report the value of a statistical robbery equal to R\$ 9,861.61\footnote{The authors use 2011 data from 4,030 households in the city of Fortaleza-CE.}, respectively. In two years, this policing assignment generated R\$ 4.31 million (US\$ 2.3 million) to improve public safety. Besides, R\$ 13.74 million (US\$ 7.4 million) was collected in fines distributed to drivers, and there are uncovered potential benefits of reducing traffic fatalities\footnote{\citeonline{deangelo2014life} finds that the sharp decrease in the number of highway troopers in Oregon due to a mass layoff increased the number of traffic injuries and fatalities.} through drunk driving crackdowns.
			
			Costs are mainly related to salaries and equipment. An average intervention consists of two vehicles and six policemen, and on any given day, there are five blitzes. Individual yearly salaries are around R\$ 50,000, and a police vehicle costs R\$ 150,000,000. Hence, 30 officers and 10 vehicles are needed to perform these daily policing assignments over the year, assuming that the same cops do not assemble more than one blitz daily. In these circumstances, the total costs for two years of interventions were approximately R\$ 6 million (US\$ 3.2 million), much lower than the revenue generated by traffic tickets.
			
			\section{Final Remarks}\label{SECTION:FinalRemarks}
			
			The present study provides robust evidence of the causal effects of place-based police interventions on violent crimes. Although there is a large body of literature showing that more police reduce crime, there is not much information on what cops should do on the streets to generate this effect. We evaluated a typical police crackdown in Brazilian cities: the allocation of blitzes. This kind of policing tactic generates deterrence by being highly visible in a street segment for a short period (30 minutes to 8 hours) and quasi-random in space-time. The intermittent design of the assignment produces uncertainty that might be useful in minimizing the temporal and spatial displacement of the crime. 
			
			We acquired high-quality, large, and fine-grained georeferenced data about criminal activities, blitzes operations, use of police input, people stopped and searched vehicles and apprehended arms. Such unprecedented micro \textit{Big Data} (a balanced panel with 7,388,808 rows) allowed us to take advantage of the high spatiotemporal resolution of the data (for both the location of blitzes and the spatial distribution of crime) to make comparisons of small intervention areas and to show that an average police crackdown causes a 35\% decrease in violent crime occurrences.
			
			The negative correlations between the time spent in a blitz, the number of vehicles selected for inspections, and the number of seized vehicles remain in accordance with well-known police experience about diminishing returns to this type of crackdown which agrees with the idea of initial deterrence decay. After a couple of hours (6 hours is approximately the maximum), drivers with potential traffic violations might be informed of the police intervention and avoid certain streets. This evidence itself has considerable efficient implications for police resource allocation.
			
			Our estimated direct and indirect (spatial and temporal) effects of local police interventions show that blitzes have a meaningful and statistically significant direct effect on violent crime occurrences, and an additional hour spent on this policing assignment causes an average decrease of 21\% in daily crime counts in cells during a given 6-hour period of day. 
			
			A range of different spatial weight matrices is proposed to assess the presence of spatial effects and three different use rings of 500-1500 meters in radius to determine catchment areas, placing higher weights on closer neighbors. In the preferred specification using an inverse distance spatial weight matrix with a 1 km cut-off point, an average decrease of 0.3\% in violent crime occurrences in each cell followed by an additional hour of policing in its neighboring region, which is not statistically different from zero, however. All in all, there is no evidence of spatial effects across a wide range of weight matrices: neither a diffusion of benefits of blitzes to nearby areas nor crime displacement to nearby cells.
			
			As to lasting effects, we could not reject the null of no joint significance of lags up to 16. Hence, we do not find residual deterrence or temporal displacement associated with blitzes. We observe neither spatial nor temporal displacement of crime, and direct effects are statistically indistinguishable from total effects.
			
			The classical distinction between deterrence and incapacitation is dealt with by means of a regression approach. Our estimates point to the former as the main driver of crime reductions. Lastly, the cost-benefit analysis of the blitzes shows that, in two years, these blitzes generated R\$ 4.31 million (US\$ 2.3 million) in improving public safety. In addition, R\$ 13.74 million (US\$ 7.4 million) was collected in fines distributed to drivers, and there are potential benefits uncovered to reduce traffic fatalities in drunk driving crackdowns. The total costs for two years of interventions were about R\$ 6 million (US\$ 3.2 million), much lower than the benefits.
			
			Our analysis can be further improved in some directions. However, we describe four possibilities believed to be feasible, more urgently needed, and achievable in the short to medium term. The first is related to the need to consider the ``excess of zeros'' characteristic of our sample, a key econometric issue. The second has to do with including a more nuanced structural econometric modeling to shed light on the workings and impacts of giving police officers discretion. The third is about facing the computational challenge brought about by our available \textit{Big Data} using high-performance computing; and the fourth proposes incorporating the multitreatment brought about by the existence of fixed and mobile blitzes.
			
			The (very) high spatiotemporal resolution of our blitzes data provided a unique opportunity to assess the causal effects of local police interventions, but also presented challenges we have so far neglected. In fact, the spatial distribution of our dependent variable (counts of crime in each of the 2,562 hexagons) shows a pattern of many zero occurrences. Although in 2012 and 2013 there were 644 and 633 hexagons that zero crimes (roughly 25\% of zeroes); when considering the time disaggregation employed by our analysis (4 periods per day, for each of the 365 days x 2 years)  then 99\% of the outcomes are zero, the setup is one of ``excess of zeros'' or ``zero inflation''. 
			
			Coupled with the fact that we have a complex spatio-temporal count \textit{Big Data} structure that nests geostatistical point-referenced data with areal data; a large array of theoretical, applied, and computational issues arises, see \citeonline{Young2022-gt}. In fact, even if we abstract from the time dimension for a while, modeling large zero-inflated spatial count datasets remains a key challenge from both computational and modeling standpoints because: i) they are overparameterized; ii) model fitting can be computationally burdensome due to large matrix operations; and iii) the spatial processes may potentially exhibit complex spatial dependence characteristics such as non-stationarity and anisotropy; see \citeonline{Lee2024-kg}.
			
			So, not considering the ``zero-inflated'' feature of our data is a source of misspecification with likely impacts on the estimated results. Hence, we see as a necessary priority tracking that misspecification. A possible starting point would be the Bayesian approach advocated by \citeonline{He2024-np}. They chose a zero-inflated negative binomial (ZINB) model, since it is a popular choice for modeling ``zero-inflated'' data, as it gives more reliable parameter estimates when non-zero counts are overdispersed compared to other models such as the ``zero-inflated'' Poisson model. In addition, their Bayesian approach produces tractable inference implemented using software such as R (\citeonline{Rsoftware}). Clearly, a ZINB specification and estimation will result in more robust and trustable estimatives. In addition, \citeonline{Cardot2021-ds} is a very recent possibility. They consider spatial spillover effects and approach them with nonparametric techniques.
			
			As police officers have a great deal of consequential discretion in any given police-citizen encounter, scholars in criminology have pointed out the importance of research on factors influencing that discretion and accompanying impacts in situations such as the use of force, minority groups bias on stop-and-frisk operations, on-duty killings, etc. See the seminal article of \citeonline{ Goldstein1963-xg} as well \citeonline{Alpert2005-du}, \citeonline{Cordner2014-gm}, \citeonline{De_Buck2024-xx}, and \citeonline{Turner2024-fw}. Economists, confronted with a large empirical and factual body of evidence, have been dealing with police discretion indirectly by focusing on race (usually called racial profiling) and the use of lethal force; see \citeonline{ Anwar2006-jy}, \citeonline{ Fryer2019-wq}, and \citeonline{ Hubert2023-cc}. Neither criminologists nor economists have closed the debate, although they provided many insights and policy recommendations.
			
			The diminishing returns to the duration of the blitzes we found (see Figure \ref{FIGURE:OutputBlitzes}) and Table \ref{TABLE:EstimationDiminishingReturns}) may hide important clues about how police officers behave. This is so not only because of the empirical evidence of a blitz duration distribution; more important, this is an outcome of the discretion police officers have. What our evidence suggests is a subtle role discretion could play in achieving an efficient allocation of police resources, something that warrants further research. After receiving the mission to set up a blitz, the commander faces a complex flux of information and specifies whether or not to stop the current blitz for each period as a function of observed and unobserved state variables (number and type of vehicles passing by, stopped and searched, the many materials apprehended, weather, physical and emotional condition of the troop, and so on). The strategy set in motion resembles an optimal stopping rule that can be derived from the solution to a stochastic dynamic programming problem in the same spirit as the seminal \citeonline{Rust1987-yv} paper and the extensive literature surveyed by \citeonline{Low2017-yc} and \citeonline{Galiani2021-sf} on structural econometrics. The central payoff of a structural model of the duration of blitzes cannot be overemphasized, as it has the potential to identify mechanisms that determine outcomes, analyze counterfactual policies, quantifying impacts as well as effects. We believe that extension could decisively enhance our knowledge of police discretion and its impacts. 
			
			Although economists have previously made little use of high-performance computing (HPC)\footnote{Custom-built supercomputers or groups of individual computers called clusters, run on-premises, in the cloud, or as a hybrid of both; either by means of CPUs or GPUs.}, some have advocated the need for powerful computing tools since at least \citeonline{Amman1989-pe}, \citeonline{Zenios1999-ox}, \citeonline{Doornik2002-lt}, \citeonline{Judd2006-lb}, and \citeonline{Aldrich2014-ln}. In fact, with the increase in size and complexity of spatio-temporal datasets, traditional methods for performing statistical and econometric analysis lag the real-time demands for decision support systems backed by scientific modeling. This is especially critical within our blitz \textit{Big Data} context, as we have discussed before. Not only there are many computational challenges (\citeonline{Young2022-gt}) but, according to \citeonline{Ajayakumar2020-kb}, there is no software available that could be scalable for large crime datasets. Hence, to solve the challenges related to \textit{Big Data}  and to support policy decision making, a high-performance computing approach is a viable and much needed alternative.
			
			Our choice of computational power can be seen in at least two directions. First, our empirical evidence is a massive geospatial \textit{Big Data} (see, \citeonline{Li2020-lh}) comprised of a lot of observed and unobserved heterogeneity (see \citeonline{Browning2010-av}) that needs to be modeled, be it spatial, individual (tastes or preferences), or technological. For example, if we generalize our econometric modeling to specify a more nuanced weight matrix \textbf{W}, or to include more observed dependent variables, or to suppose heterogeneous impacts of the blitzes, or to depart from the SLX model; the resulting heterogeneity translates into a high-dimensional system subject to the ``curse of dimensionality'' (\citeonline{Eftekhari2017-pj}), and a viable solution is HPC. Second, our suggested structural econometric model of police discretion results in a nonlinear, and very high-dimensional dynamic stochastic economic model that requires HPC. For example, \citeonline{Brumm2017-db} implementation of a fully hybrid parallel solution, combining distributed and shared memory parallelization paradigms; or the more recent \citeonline{Eftekhari2022-jy} iteration framework that approximates economic policy functions using an adaptive, high-dimensional model representation scheme, combined with adaptive sparse grids to address the ubiquitous challenge of the curse of dimensionality are two interesting avenues for exploration.
			
			We have not considered the multiple treatment feature of our set-up represented by the existence of two types of blitzes: fixed and mobile. Except for acknowledging the existence of different blitzes (see Figure \ref{FIGURE:TypeBlitzes}), this heterogeneity is used again only once to shed light on the likely mechanism impacting crime, either deterrence or incapacitation. However, estimating the impacts by explicitly considering these two blitz models is an interesting avenue of research improvement.  With three treatment options (that is, the hexagon does not receive a blitz, or it receives a ``fixed'' blitz, or a ``mobile'' blitz) or more, the estimation of causal effects requires additional identification assumptions, estimation techniques, and conceptualizations; as the increasing number and sophistication of the econometric literature (\citeonline{Frolich2004-kb}), \citeonline{Lee2005-vo} and \citeonline{AtheyImbens2017}) as well as the statistical (\citeonline{Lopez2017-rf}) attests.
			
			There are 4 main strategies for analyzing multiple treatments (for binary treatments as well): i) selection on observables, see \citeonline{Lechner2001-zb}, ii) instrumental variable, see \citeonline{Bhuller2024-cg}, iii) difference-in-differences, see \citeonline{Fricke2017-lk}, and iv) regression discontinuity design, see \citeonline{Gilraine2020-dr}, and \citeonline{Caetano2023-ce}. Note that the combination of our ``excess of zero'' issue (for instances, these zeros are ``structural zeros'' where the count response will always be zero; or ``random zeros'' where there could have been an event, but there was none; see \citeonline{Gilraine2020-dr}) with the multiple treatment dimension implies a considerable challenge, yet to be conceived. At this juncture, we can only conjecture the best way to follow and suggest a selection-on-observables approach as the best way to deal with our multiple treatament ``excess of zero'' model.

			

			

			
			\renewcommand{\bibname}{References}
			\bibliography{refs}
			

		\end{document}